\documentclass[12pt%
 preprint,
 amsmath,amssymb,
 aps,
]{revtex4-2}

\usepackage{xcolor}
\usepackage{graphicx}
\usepackage{dcolumn}
\usepackage{bm}
\usepackage{hyperref}
\usepackage{float}
\usepackage{ulem}
\usepackage{geometry}
\usepackage{comment}
\usepackage{subfigure}

\newcommand{\sinc}{\mathrm{sinc}}

\newcommand{\sincxb}{\sinc\left(\frac{\overline{c}_\psi l_x}{2}\right)}

\newcommand{\ovl}{\overline}
\newcommand{\xb}{\overline{x}}

\newcommand{\bs}{\boldsymbol}

\newcommand{\wt}{\widetilde}

\newcommand{\xt}{\Tilde{x}}
\newcommand{\yt}{\Tilde{y}}
\newcommand{\zt}{\Tilde{z}}

\newcommand{\be}{\begin{equation}}
\newcommand{\ee}{\end{equation}}
\newcommand{\ba}{\begin{eqnarray}}
\newcommand{\ea}{\end{eqnarray}}

\newcommand{\bse}{\begin{subequations}}
\newcommand{\ese}{\end{subequations}}

\newcommand{\pd}{\partial}

\begin{document}

\preprint{APS/123-QED}

\title{Preferential orientation of slender elastic floaters in gravity waves}

\author{Wietze Herreman}
\email{wietze.herreman@universite-paris-saclay.fr}

\author{Basile Dhote}

\author{Fr\'ed\'eric Moisy}%
\affiliation{Université Paris-Saclay, CNRS, Laboratoire FAST, 91405 Orsay, France \\}%


\date{\today}

\begin{abstract}
Slender floaters drifting in propagating gravity waves slowly rotate towards a preferential state of orientation with respect to the angle of incidence. This angular drift arises from a wave-induced, second order mean yaw moment. We develop a diffractionless, hydro-elastic theory to compute this mean yaw moment for a thin, flexible structure  whose width and thickness are small compared with the wavelength. For floater lengths smaller than half the wavelength, we derive a simple, predictive criterion for the preferred orientation: Soft, short and heavy floaters prefer the longitudinal state, while stiff, long and light floaters prefer the transverse state.  For floaters longer than the wavelength, the orientational dynamics become more intricate and may exhibit multiple equilibrium states.  We discuss the implications of the model for flexible floating structures such as pontoons and inflatable structures.
\end{abstract}


\maketitle

\section{Introduction}

In many naval engineering problems, it is necessary to predict the response of floating structures to gravity waves~\cite{newman2018marine,kim2008nonlinear,molin_2023}.
While a large part of work focuses on rigid bodies, numerous applications require to model the wave-induced motion of deformable and flexible floating structures. This is the subject of hydro-elasticity \cite{bishop_1979}. Typical examples include thin plates or sheets in waves \cite{montiel_hydroelastic_2013-1,montiel_hydroelastic_2013,wang_2015}, ice-floes \cite{meylan_response_1994,squire_2020},  modular pontoons \cite{watanabe2004hydroelastic}, floating solar farms \cite{oceansun}, offshore floating airports \cite{isobe1999research}, flexible wave-energy converters  \cite{mozaf2025performance} and even ship-like structures \cite{malenica2013homer}. An early review on pontoon-type hydroelastic structures can be found in~\citet{watanabe2004hydroelastic} and more recent advances are discussed in~\citet{zhang_2022} and \citet{TAVAKOLI2025}.

\begin{figure}
    \centering
    \includegraphics[width = 0.6\textwidth]{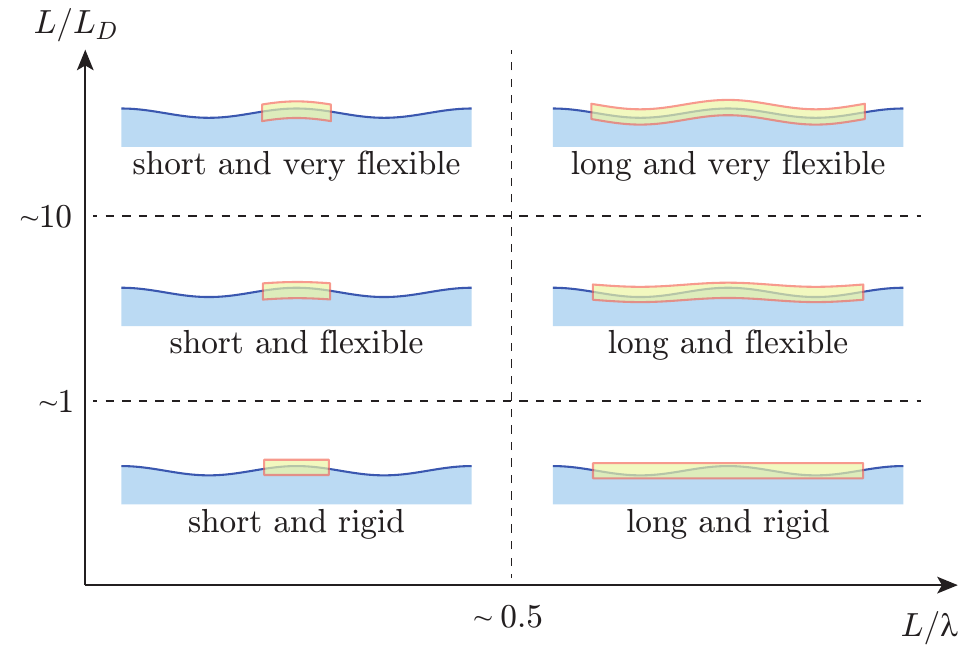}
    \caption{Typical deformation of a thin elastic floating structure in waves as a function of the ratios $L/\lambda$ (floater size/ wavelength) and $L/L_D$ (floater size/ flexural length).} 
    \label{FIGD}
\end{figure}

Many hydroelastic applications involve thin floating plates, whose bending deformation can be described using the Kirchhoff–Love plate equation~\cite{timoshenko1959plates,landau1986elasticity}. For thin floating plates, the flexural length $L_D =  (  {D/\rho g}  )^{1/4}$
sets an important length-scale, with $D$ the bending modulus of the plate, $\rho$ the density of water and $g$ gravity. In figure \ref{FIGD}, we illustrate the typical instantaneous deformation of a plate with length $L$ in waves with wavelength $\lambda$.  For high ratios $L/ L_D$ the plate is very flexible and it locally adapts to the wave. For low ratios $L/ L_D$, the structure weakly deforms and shows a spatially varying submersion depth. 

While it is often sufficient to determine the linear, first order response of the structure to the incoming wave, higher-order corrections can also be important for applications. 
At second order, waves induce both a mean drift force and a mean yaw moment on the structure \cite{newman1967drift}, which may have implications for the design and performance of mooring systems. For rigid floating structures, the calculation of the second order mean load has become standard, with existing software packages such as Hydrostar \cite{veritas2016hydrostar}. For flexible floating structures, the calculation of the second order mean load is less standard. An early study is that of \citet{kashiwagi1998new}, and other examples are mentioned in the review of \citet{watanabe2004hydroelastic}. More recently, \citet{miao2019analysis} investigated wave-induced mean loads on a flexible pontoon.

\begin{figure}[t]
    \centering
    \includegraphics[width = \linewidth]{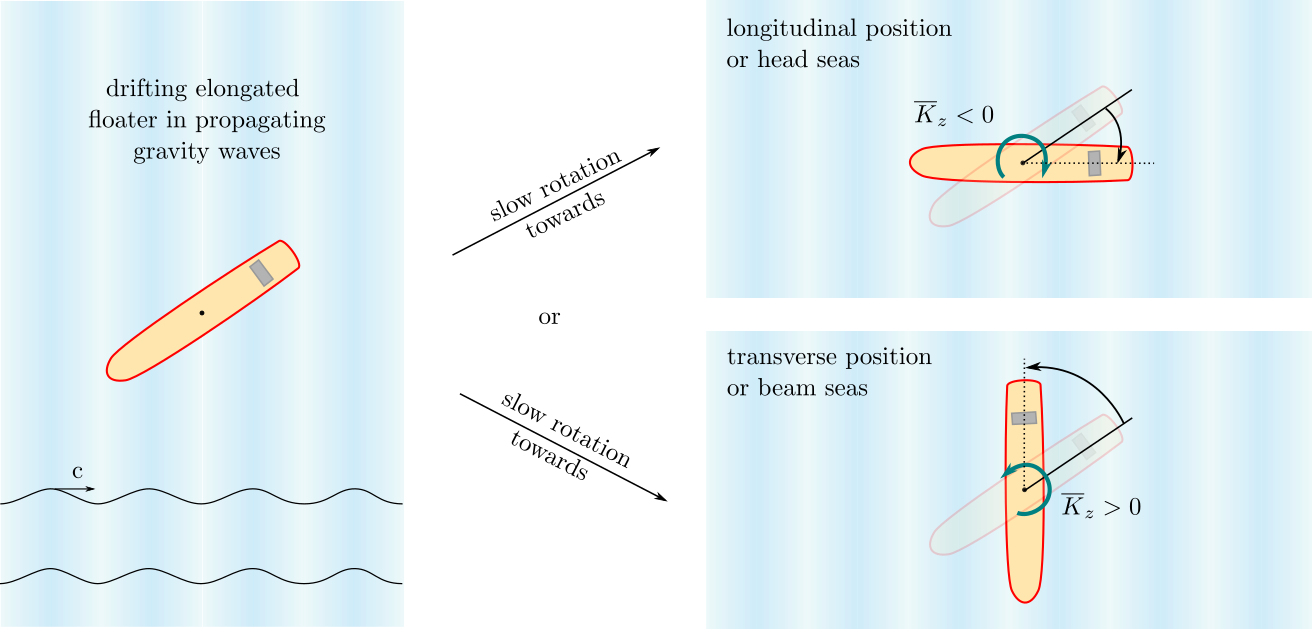}
    \caption{Elongated, freely drifting structures  placed in propagating gravity waves slowly rotate towards a preferential orientation that can be either longitudinal (head seas), transverse (beam seas) or somewhere in between. }
    \label{sketch_LT}
\end{figure}

For non-moored floating structures, the  mean force and moment induce a drift in position and orientation. Wave-induced drift in position  has been extensively studied for small solid bodies~\cite{van_den_bremer_stokes_2018,monismith_stokes_2020} (commonly referred to as Stokes drift in this context) and also for thin elastic sheets or plates \cite{law_wave-induced_1999,wong_wave-induced_2003,christensen_wave-induced_2005,law_observations_2007,wang_2015,kostikov_2021}. Wave-induced angular drift has been much less studied but it plays an important role in seakeeping \cite{skejic2008unified}. The mean yaw moment permanently acts on the course of boats, an effect that can be felt when maneuvering kayaks in wavy lakes or seas.  As illustrated in  Figure~\ref{sketch_LT},  the mean yaw moment can rotate slender floaters towards preferential states of orientation. We may naturally anticipate the longitudinal (\textit{head seas}, with the long axis aligned with the direction of incidence) and the transverse (\textit{beam seas}, with the long axis perpendicular to the direction of incidence) orientations as equilibrium configurations, but intermediate equilibria can also arise.  The preferential orientation of slender floaters in waves was first observed a century ago by \citet{suyehiro1921yawing} using small, solid boat models in a wave tank, and a first theoretical explanation was proposed by \citet{newman1967drift}. More recently, \citet{wong_wave-induced_2003} and  \citet{boulluec_steady_2008} observed angular drift in experiments involving, respectively, freely drifting elliptical plates and container ship models.

In recent work, we have described the preferential orientation mechanism using a diffractionless theory \cite{herreman2024preferential,dhote2025flexible}. In this framework, the incident wave field is assumed to remain unaltered by the presence of the floater, which amounts to neglecting diffraction and radiation effects (Froude-Krylov approximation). The theory applies to both rigid \cite{herreman2024preferential} and perfectly flexible floaters \cite{dhote2025flexible} smaller than the wavelength, and received experimental validation. The present work extends this diffractionless approach to the hydroelastic regime for thin slender floating plates in waves. By introducing bending rigidity in the model, we can connect the solid and perfectly flexible limits that were previously described. Our model is most predictive for slender floaters whose three characteristic dimensions remain small compared with the wavelength:~{\it wavelength $\gg$ length $\gg$  width $\gg$ height}. In this limit, we find that preferential orientation is either longitudinal or transverse, depending on how the non-dimensional number $F$ compares to a critical value $F_c$:
\bse \label{prediction}
\be \label{Fccrit}
 F = \frac{kL_x^2}{\bar{h}} \quad, \quad \left \{ \begin{array}{rcl}
F< F_c & , & \text{longitudinal} \\
F > F_c & , & \text{transverse} 
\end{array} \right.
\ee
where  $k$ is the incoming wavenumber, $L_x$ the floater's length and $\overline{h}$ the equilibrium submersion depth. The critical value $F_c$ depends on the shape and on the rigidity of the floater. For solid rectangular parallelepipeds, we obtain $F_c=60$, in close agreement with experimental observations \cite{herreman2024preferential}. For very flexible strips, we found that the preferential orientation is always longitudinal \cite{dhote2025flexible}, meaning that $F_c \rightarrow +\infty$ in that limit.  In the present article, we show that, for slender elastic plates, the critical value is
\be \label{Fcdef_intro}
F_c \approx  60 + \frac{5}{42} \left( \frac{L_x}{L_D} \right)^4.
\ee
\ese
This formula correctly reproduces the rigid and flexible limits, $F_c=60$  and $F_c \rightarrow + \infty$, respectively. The origin of the variation of $F_c$ with $L_x/L_D$ is the non-uniform submersion depth along the long axis. As shown in Refs.~\cite{herreman2024preferential} and \cite{dhote2025flexible}, spatial variation in submersion has a strong impact on a part of the mean yaw moment that favors the transverse orientation. As can be seen in figure \ref{FIGD}, immersion depth is indeed strongly affected by the bending rigidity: it is maximally varying for solid floaters, constant for perfectly flexible floaters, and somewhere in between for elastic floaters. 

The Froude–Krylov assumption underlying our model usually applies to floaters that are small compared with the wavelength in all three dimensions. For longer floaters, the influence of the floater’s motion on the incident wave (diffraction and radiation effects) cannot, in principle, be neglected. Nonetheless, we provide an argument showing that this diffractionless approach remains valid for floaters of arbitrary length, {\it as long as the remaining two dimensions (width and thickness) remain small}. This argument relies on the observation, first made in Ref.~\cite{herreman2024preferential} and detailed here in Appendix \ref{sec:appendix}, that our diffractionless, near-field theory exactly recovers Newman's diffraction-based far-field formula \cite{newman1967drift} of the mean yaw moment on a slender rigid body of arbitrary length. If a diffractionless near-field approach yields the same result as a diffraction-based far-field approach, it implies that {\it diffraction corrections of pressure are negligible in the near field}. While this reasoning strictly applies to rigid floaters, it can be safely extended to flexible ones, as flexible floaters produce even less diffraction. This supports the use of a diffractionless hydro-elastic theory for the case of slender flexible floaters with arbitrary length and two short directions. 

The theoretical predictions for floaters that are not short with respect to the wavelength are generally not as simple as that of Eq.~\eqref{prediction}: longer floaters can also stabilize at intermediate equilibrium angles and there can even be multiple equilibrium positions, so preferential orientation may be sensitive to initial conditions.  By comparing the arbitrary-length theory to the short limit, we can specify the domain of validity of the simple prediction \eqref{prediction}. 

The article is structured as follows.  In section \ref{sec:theory}, we define our hydro-elastic model and we calculate the second order mean yaw moment on slender elastic floaters. We discuss how the mean yaw moment varies with angle of incidence, floater length and bending rigidity.  The arbitrary length theory is then simplified in the short limit where we find the simple criterium \eqref{prediction}. In section  \ref{sec:applications}, we apply our theory to typical structures such as flexible pontoons, inflatable structures, and we also discuss the design of experiments using polyethylene foam structures that could be carried out in medium-scale wave flumes. Section \ref{sec:conclusion} presents the conclusions.

\section{Froude-Krylov model for the mean yaw moment on slender elastic floaters in waves}
\label{sec:theory}

\begin{figure}
    \centering
    \includegraphics[width = \linewidth]{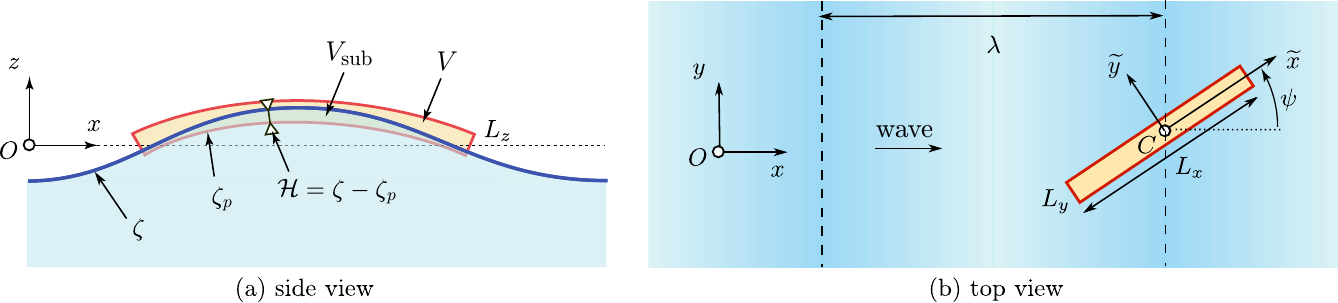}
    \caption{(a) A flexible strip with center $C$ and dimensions $L_x, L_y, L_z$ is being displaced by a propagating gravity wave. In the side-view (a) we imagine $\zeta_p $, the bottomline of the strip to be different from $\zeta$, the water surface. The local submersion depth $\mathcal{H} = \zeta - \zeta_p $ depends on the position of the floater in the wave and on the bending rigidity. In the top-view (b), the yaw angle $\psi$ is the angle between the strip long axis $\wt{x}$ and the direction of wave propagation $x$.}
    \label{FIG1}
\end{figure}

We propose a diffractionless, Froude-Krylov model to calculate the wave-induced, second order mean yaw moment on slender elastic floaters with arbitrary length and two short directions. The role of diffraction in this problem is discussed in detail in Appendix \ref{sec:appendix}. 
Our methodology follows that of \citet{herreman2024preferential} and \citet{dhote2025flexible}, complemented here by a Kirchhoff–Love model to describe the bending of the thin plate.

\subsection{Equations of motion \& simplifying assumptions}

The system is sketched  in  Figure \ref{FIG1}. The incoming wave is idealized as a linear inviscid potential wave in infinitely deep water. In the reference frame $(O,x,y,z)$, the surface elevation and velocity potential are
\begin{eqnarray}
    \zeta (x,t) & = & a \sin(kx - \omega t), \hspace{.5 cm} {\phi} (x,z,t) = - \frac{a \omega}{k} e^{kz} \cos(kx - \omega t),  \hspace{.5 cm} p = p_0 - \rho g z - \rho \partial_t\phi . \label{eqflow}
\end{eqnarray}
We denote $a$ the wave amplitude, $k$ the wavenumber, $\omega = \sqrt{gk}$ the frequency, $g$ the gravitational acceleration, $p_0$ the atmospheric pressure and $\rho$ the fluid density. The fluid velocity is $\bs{u} = \bs{\nabla} \phi$.  We suppose a small slope, meaning that
\be
\epsilon = k a \ll 1.
\ee
Second order  corrections ($\sim a^2$) to the wave were given in \citet{herreman2024preferential}. They create a weak surface elevation and a stationary pressure modification, but the flow remains unaffected. As explained below, the first order, linear approximation of flow is sufficient to calculate the second order mean yaw moment with our approach. 

The floater is a thin rectangular plate with density $\rho_p = \beta \rho$, with $\beta < 1$ the density ratio. We denote its length $L_x$, width $L_y$ and height $L_z$ and we suppose a scale separation 
\begin{eqnarray}
 L_x  \gg L_y \gg L_z \gg \text{capillary length}.
\end{eqnarray}
The capillary length, typically a few mm in water, is much smaller than all dimensions of floating structures that relate to maritime applications. Hence, we ignore capillary effects in what follows. The wavelength $\lambda$ is supposed long with respect to the width and height and depending on how it compares to the length, we make a distinction between long and short floaters:  
\be
\lambda  \gg L_y \gg L_z \quad , \quad 
\left \{ 
\begin{array}{rcl}
L_x \gg \lambda & , & \text{long floater}  \\ 
L_x \ll \lambda & , & \text{short floater} 
\end{array} \right . .
\ee
To describe the floater motion, we introduce a second reference frame $(C,\wt{x},\wt{y},\wt{z})$. The point $C$ is the projection of the center of the floater on the $z=0$ plane. Relative to $O$, this point $C$ has coordinates $x_c (t)$ and $y_c$. In our model, $y_c$ remains constant and it is arbitrary. We introduce the {yaw} angle $\psi(t)$ that measures the orientation of the long $\wt{x}$-axis with respect to the $x$-axis of wave propagation (see Figure  \ref{FIG1}).  

With $\epsilon \ll 1$, the wave-induced deformation of the plate is weak. We can write the approximate, laboratory frame coordinates of any point $x_p,y_p,z_p$ in the thin plate as
\be
\left \{ \begin{array}{rcl}
     x_p&=& x_c (t) + \xt \cos  \psi (t) - \wt{y} \sin  \psi (t)   \\ 
     y_p & =& y_c + \xt \sin \psi (t) + \yt \cos \psi (t)  \\
     z_p & =&  \zeta_p (\wt{x},\wt{y},t) + \widetilde{z}
\end{array} \right .  . \label{tf_zt}
\ee
Varying $\wt{x} \in [-L_x/2,L_x/2]$, $\wt{y} \in [-L_y/2,L_y/2]$, $\wt{z} \in [0,L_z]$ we cover the floater volume $V$. The function $\zeta_p (\wt{x},\wt{y},t)$ represents the bottom of the plate and in the linear regime, it satisfies the Kirchhoff-Love equation~\cite{timoshenko1959plates,landau1986elasticity}, here 
\be \label{KL2D}
D \wt{\nabla}^2  \wt{\nabla}^2 \zeta_p + \underbrace{\rho_p L_z \frac{\partial^2  \zeta_p}{\partial t^2}}_{\text{negligible}}  = p |_{z=\zeta_p} - p_0 - \rho_p L_z g .
\ee
The bending rigidity or bending modulus of the thin plate is defined as
\be
D = \frac{E L_z^3 }{12(1- \nu^2)}
\ee
with $E$ the Young modulus and $\nu$ the Poisson ratio of the strip. The boundaries of the plate are free, which imposes zero bending moment and zero shear stress at the rim. These conditions can be expressed as :
\be
\left . \frac{\pd^2 \zeta_p}{\pd \wt{x}^2} + \nu \frac{\pd^2 \zeta_p}{\pd \wt{y}^2} \right |_{\wt{x} =\pm L_x/2 } = 0  \quad , \quad \left .  \frac{\pd^3 \zeta_p}{\pd \wt{x}^3} + (2- \nu) \frac{\pd^3 \zeta_p}{\pd \wt{x} \pd \wt{y}^2} \right |_{\wt{x} =\pm L_x/2 }  = 0 
\ee
and similar at $\wt{y} = \pm L_y/2$, inverting $\wt{x} \leftrightarrow \wt{y}$ (see for example Ref.~\cite{kashiwagi1999time}). Using the gravity wave dispersion relation, we can show that the plate inertia is negligible with respect to the pressure term under the assumption $k L_z \ll 1$, as indicated by the brace in  \eqref{KL2D}. In the right hand side of the Kirchoff-Love equation, we find weight and pressure terms. In our Froude-Krylov approach, we can use the pressure field defined above and this gives
\be
p |_{z=\zeta_p} - p_0 \approx - \rho g \zeta_p  - \rho \pd_t \phi |_{z=\zeta_p} .
\ee
This relation can further be simplified. To explain how, we first recognise that the equilibrium position of the plate is $\zeta_p = - \beta L_z$ in the absence of wave. For all values of the bending modulus $D$, the plate is submerged at depth $\beta L_z$ and it remains flat. In waves, the plate deforms and for our calculation of the second order mean yaw moment, we only need the first order plate deformation ($\sim a$). To get this first order deformation, it is sufficient to evaluate the dynamic pressure term at the equilibrium depth $z=-\beta L_z$. But since we also assume that $k\beta L_z \ll 1$ we can further simplify this and evaluate the dynamical pressure at depth $z=0$. In summary, we can simplify the dynamical pressure term as
\be
\rho \pd_t \phi |_{z=\zeta_p}  \approx \rho \pd_t \phi |_{z=-\beta L_z } \approx \rho \pd_t \phi |_{z= 0 }  .
\ee
With the expression of the wave-potential defined in Eq.~\eqref{eqflow}, and replacing $x=x_c (t) + \xt \cos  \psi (t) - \wt{y} \sin  \psi (t)$, we have reduced the plate deformation problem to that of solving
\be \label{KL2D_red}
D \wt{\nabla}^2  \wt{\nabla}^2 \zeta_p + \rho g \zeta_p  =   \rho g \left( -\beta L_z + a \sin (kx_c - \omega t + k \xt \cos  \psi - k\wt{y} \sin  \psi ) \right  ) .
\ee
Let us briefly analyse this equation. In the left hand side, both terms are of same order of magnitude when deformations occur on the flexural length-scale $L_D =( D/ \rho g )^{1/4}$. The right hand side is forcing a deformation on the scale of the wave-length $\lambda$. Qualitatively, we expect a nearly rigid motion of the plate when $L_D \gg L_x, L_y$, and deformations at the scale of $\lambda$  with elastic boundary layers of size $L_D$  when $L_D \ll L_x, L_y$. 

We now further simplify our model by assuming the scale separation 
\be
L_D \gg L_y.
\ee
In this limit, we can neglect the bending deformation in the $\wt{y}$-direction and decompose $\zeta_p$ into a bending deformation that is $\wt{x}$-independent and a twisting deformation that varies linearly with $\wt{y}$ : 
\be \label{decomp}
\zeta_p  (\wt{x},\wt{y},t) \approx \underbrace{  f (\wt{x},t) }_{\text{bending}} + \underbrace{ \wt{y} \,  \varphi  (\wt{x},t) }_{\text{twisting}}
\ee
The function $\varphi  (\wt{x},t)$ represents a local angle of twist or roll. At the end, we will find that twist deformations are negligible in the mean yaw moment when $L_x \gg L_y$, but they are retained here for completeness and to avoid any ambiguity. We inject this decomposition \eqref{decomp} into \eqref{KL2D_red} and replace the right hand side with a first order Taylor series in the $\yt$-direction \ba
&& \sin (k x_c- \omega t + k\xt c_\psi   - k\yt s_\psi   )  \nonumber \\
&& \quad  \quad  \approx  \sin (k x_c  - \omega t + k\xt c_\psi )  -  k \yt s_\psi \,  \cos ( k x_c - \omega t + k\xt c_\psi  ).
\ea
We denote $c_\psi = \cos \psi$ and $s_\psi =\sin \psi$. This leads to a pair of equations for $f (\wt{x},t)$ and $\varphi  (\wt{x},t) $:
\bse \label{KL2D_red2}
\ba 
\frac{D}{\rho g} \frac{\pd^4 f }{\pd \xt^4} +  f  &=&- \beta L_z +  a  \sin (k x_c  - \omega t +k \xt c_\psi ) \\
\frac{D}{\rho g} \frac{\pd^4 \varphi }{\pd \xt^4} +  \varphi  &=&  -  k a s_\psi  \cos (kx_c - \omega t + k\xt c_\psi)  .
\ea
\ese
The free-plate boundary conditions reduce to
\be  
\left . \frac{\pd^2 f}{\pd \wt{x}^2} \right |_{\wt{x} = \pm L_x /2} =   \left . \frac{\pd^2 \varphi}{\pd \wt{x}^2} \right |_{\wt{x} = \pm L_x /2} = 0 \quad , \quad \left . \frac{\pd^3 f}{\pd \wt{x}^3} \right |_{\wt{x} = \pm L_x /2} = \left . \frac{\pd^3 \varphi}{\pd \wt{x}^3} \right |_{\wt{x} = \pm L_x /2} = 0 .
\ee
Both problems can be solved analytically and with $f$ and $\varphi$ known, we can calculate the local submersion depth as the difference $\mathcal{H} = \zeta - \zeta_p  $. Using first order Taylor series in $\wt{y}$, we can also isolate two parts there :
\ba 
\mathcal{H} (\xt,\yt,t) & \approx& h(\xt,t)  + \yt\,  \alpha (\xt,t)    \label{immersion}  \\ 
&\approx& \underbrace{ a  \sin (k x_c - \omega t + k\xt c_\psi ) - f }_{\text{bending}}  + \underbrace{ \yt \left (  -  k a s_\psi  \cos (kx_c  - \omega t + k \xt c_\psi )  - \varphi \right ) }_{\text{twisting}} \nonumber
\ea
We denote $h (\xt,t)$ the local variation in submersion due to bending along $\wt{x}$. The part  $\yt\,  \alpha (\xt,t)$ is due to twisting. 
The local submersion is an important quantity in our theory and we implicitly assume that the floater is never fully submerged nor de-wetted, $0 \leq \mathcal{H}  \leq L_z$. In practice this sets a non-trivial limitation to the maximal incoming wave-magnitude. 

With the equations for the first order plate deformation specified, we turn to the evolution equations for $x_c(t)$ and $\psi(t)$. These are given by Newton's law ($x$-component, in the inertial laboratory frame) and the angular momentum theorem ($z$ or $\wt{z}$-component, in the non-inertial, floater frame):
\begin{eqnarray}  
        \frac{d}{dt}\underbrace{\left(\int_{V} \rho_p  v_x \, dV\right)}_{ =\, M\dot{x}_c} = F_x, \quad\quad \frac{d}{dt}\underbrace{\left(\int_{V}\rho_p  \, \bs{e}_z \cdot (\bs{r} - \bs{r}_c) \times \bs{v}  \, dV\right)}_{= \, I_{zz}\dot{\psi}} = K_z.
\label{Motion_eq}
\end{eqnarray}
The integrals in the left hand sides cover the total floater volume $V$ and we can simplify them using the local speed of the points of the plate, $v_x = \dot{x}_p, v_y = \dot{y}_p$, taking the time-derivative of \eqref{tf_zt}. We denote $M=\rho_p  L_x L_y  L_z$ the floater mass. The moment of inertia with respect to the vertical axis can be approximated $I_{zz} \approx M L_x^2 /12$ considering that $L_x \gg L_y$ by assumption. In the right hand side, we find the pressure force $F_x$ and moment $K_z$. In our Froude-Krylov model, we calculate them with the pressure of the incoming wave. Corrections of pressure due to diffraction and radiation are assumed small and they are ignored in our model (see discussion in appendix \ref{sec:appendix}). In practice, we have to calculate the surface integrals  
\begin{subequations} \label{FetK_1}
\begin{eqnarray} 
    F_x  & = &  -\int_{S_{\text{sub}}} (p - p_0) \, \bs{dS} \cdot \bs{e}_x  \label{Fx_1}\\
     K_z & = &   -\int_{S_\text{sub}} \left ( ( \bs{r} - \bs{r}_c ) \times (p - p_0) \bs{dS} \right ) \cdot \bs{e}_z   \label{Kz_1}  ,
\end{eqnarray}
\end{subequations}
with $p$ as in Eq.~\eqref{eqflow} and integrating over $S_{\text{sub}}$, the time-dependent, wetted part of the floater surface. In these formula, we orient the surface element $d\bs{S}$ from the floater towards the liquid which explains the minus sign. These integrals are not trivial to evaluate. As in \cite{herreman2024preferential,dhote2025flexible}, we can simplify the analytical calculation of $F_x$ and $K_z$ by rewriting them as volume integrals: 
\begin{subequations} \label{FxKz} 
\begin{eqnarray} 
    F_x  & = &   \int_{V_{\text{sub}}} \rho \Big (  \partial_t u_x    + \underbrace{(\mathbf{u}\cdot\boldsymbol{\nabla})u_x}_{= \, 0} \Big ) dV \label{Fx}\\
     K_z & = &    -\int_{V_\text{sub}}  \rho (y - y_c)  \Big (  \partial_t u_x + \underbrace{(\mathbf{u}\cdot\boldsymbol{\nabla})u_x}_{ = \,  0} \Big ) \label{Kz}   dV.
\end{eqnarray}
\end{subequations}
The volume integrals cover the {\it interior} of the submerged volume $V_{\text{sub}}$ that is delimited by $S_{\text{sub}}$ and the prolongation of the free surface $\zeta$ inside the floater. This reformulation is quite uncommon in floater-wave interaction theory and it is only possible within the Froude–Krylov approximation. The simplification of the pressure force integral goes as follows. Since $p = p_0$ on the free surface, we can write $\bs{F} = - \oint_{\delta V_{\text{sub}}} (p-p_0) d \bs{S}$, with $\delta V_{\text{sub}}$ the boundary of $V_{\text{sub}}$ . Using the divergence theorem, we rewrite this as $\bs{F} = - \int_{V_{\text{sub}}} \bs{\nabla} p \, dV $. With Euler's law, we can replace $-  \bs{\nabla} p  = \rho \pd_t \bs{u} + \rho (\bs{u} \cdot \bs{\nabla} ) \bs{u} + \rho g \bs{e}_z$ and this gives our formula. The manipulation for $K_z$ is similar but slightly more complex due to the extra factor $(y-y_c)$. As suggested by the braces in \eqref{FxKz}, the nonlinear term vanishes for a inviscid monochromatic wave: $(\mathbf{u}\cdot\boldsymbol{\nabla})u_x = 0 $ with \eqref{eqflow}. Considering that $u_x$ defined in \eqref{eqflow} is correct up to second order in wave magnitude, these formulas are correct at second order. 

We now approximate the calculation of the volume integrals, by taking into account the scale separation $L_x \gg  L_y \gg L_z$. We inject $u_x = a \omega e^{kz} \sin (kx - \omega t)$ in the integral and replace the laboratory frame coordinates with Eqs.~\eqref{tf_zt}. The top of the submerged volume is at the free surface position $z = \zeta$, so to integrate over the submerged volume $V_{\text{sub}}$, we must use the bounds $\xt \in [-L_x/2, L_x/2],  \yt \in [-L_y/2, L_y/2]$ and $\zt \in [0, \mathcal{H} (\wt{x},\wt{y},t) ]$. As the plate is much thinner than the wavelength, $kL_z \ll 1$, the integrands vary very little in the $\wt{z}$-direction and we so can approximate them by their value at $z=\zeta_p$. Integration over $\wt{z}$ then yields  
\bse \label{FxKz_surf}
\begin{eqnarray} 
    F_x & \approx & - \int_{-L_x/2}^{L_x/2} \int_{-L_y/2}^{L_y/2} \rho a \omega^2 e^{k \zeta_p}  \cos (k x_c  - \omega t + k \xt c_\psi - k \wt{y} s_\psi   )  \,  \mathcal{H}  \,  d \wt{x} \,  d \wt{y} , \\
     K_z &  \approx  &    \int_{-L_x/2}^{L_x/2} \int_{-L_y/2}^{L_y/2}   \rho a \omega^2 (\xt \sin \psi  + \yt \cos \psi   )   e^{k \zeta_p}  \cos (k x_c  - \omega t  + k \xt c_\psi - k \wt{y} s_\psi   ) \,   \mathcal{H}  \,  d \wt{x} \,  d \wt{y}  .
\end{eqnarray}
\ese
With $L_x \gg L_y$, we can further simplify these surface integrals to line integrals. To compute the mean yaw moment at second order, we only need $F_x$ at first order  and therefore we can approximate the submersion depth to $ \mathcal{H} \approx \beta L_z$ in the $F_x$-integral. We also simplify the term  $e^{k \zeta_p} \approx e^{-k \beta L_z} \approx 1 - k \beta L_z \approx 1 $ because $ k L_z \ll 1$. To simplify the integration along the middle axis $\wt{y}$, we replace the cosine with a Taylor series about $\wt{y}=0$:
\ba
&& \cos (k x_c   - \omega t  + k \xt c_\psi  - k \wt{y} s_\psi  ) \nonumber \\
&& \quad \quad \approx  \cos (k x_c   - \omega t  + k \xt c_\psi   ) +  k \wt{y} s_\psi \sin (k x_c   - \omega t  + k \xt c_\psi   ). \label{taylorcos}
\ea
After integration along $\wt{y}$, we get the following formula to calculate $F_x$, up to first order in $a$ and for slender bodies with $L_x \gg L_y$:
\be
F_x \approx -  \rho a \omega^2  \beta L_y L_z  \int_{-L_x/2}^{L_x/2}  \cos (k x_c   - \omega t  + k \xt c_\psi ) \,  d \wt{x}  + \ldots
\ee
The dots represent smaller terms that are due to the $\wt{y}$-variation and using the Taylor series along $\wt{y}$, we find that they are at least a factor $L_y^2 /L_x^2 \ll 1$ smaller. We may neglect them in what follows.  The formula for the yaw moment $K_z$ is simplified in a similar way, but needs to remain at second order in wave-magnitude. We simplify $e^{k \zeta_p} \approx 1 + k \zeta_p  \approx 1 + k f + \wt{y} \varphi $ and we replace the cosine with the Taylor series \eqref{taylorcos} and $ \mathcal{H} = h  + \yt\,  \alpha  $. Integration over $\yt$ yields 
\be
 K_z   \approx      \rho a \omega^2 L_y  \int_{-L_x/2}^{L_x/2} \xt s_\psi (1 +  k f(\wt{x},t) )   \cos (k x_c  - \omega t  + k \xt c_\psi   ) \,  h (\xt,t) \,  d \wt{x} \,  + \ldots 
\ee
Dots represent negligible terms due to $\wt{y}$-variation that are a factor $L_y^2 /L_x^2 \ll 1$ smaller. The variables $\varphi$ and $\alpha$ are absent in the leading order formulas for $F_x$ and $K_z$ and so, they do not need to be calculated. Under the assumptions $L_x \gg L_y \gg L_z$, $\lambda \gg L_y \gg L_z$  and $L_D \gg L_y$, twisting deformations do not contribute to the leading order mean yaw moment.

We collect all equations and non-dimensionalize space in units of $k^{-1}$, time in units $\omega^{-1}$, and mass in units $M$. We denote $l_{x,y,z} = kL_{x,y,z}$ and $l_D = k L_D$, the non-dimensional floater sizes and flexural length. We keep the notation $\xt$ and $t$ in this non-dimensional representation (although they actually represents the dimensional $k\xt$ and $\omega t$) and the same notations for the field variables.  
The non-dimensional problem that defines the bending deformation $f(\wt{x},t)$ is
\bse \label{nondimprob}
\be
l_D^4 \frac{\pd^4 f }{\pd \wt{x}^4}  + f = \epsilon \sin{(x_c - t + c_\psi \xt  )} - \beta l_z
\ee
with boundary conditions
\be
\left . \frac{\pd^2 f }{\pd \wt{x}^2} \right |_{\wt{x}= \pm l_x /2} =  \left . \frac{\pd^3 f}{\pd \wt{x}^3} \right |_{\wt{x} = \pm l_x /2} = 0.
\ee
With the bending deformation, we calculate the local submersion depth  
\be
h (\xt,t) =  \epsilon \sin{(x_c - t + c_\psi \xt  )}  - f (\xt,t)
\ee
The non-dimensional equations of motion for $x_c$ and $\psi$ are:
\begin{eqnarray}
\ddot{x}_c& \approx & -  \frac{\epsilon}{l_x} \int_{-l_x/2}^{l_x/2}   \cos{(x_c - t + c_\psi \xt )}   \, d\xt \label{xc_int} \\
\ddot{\psi}  &\approx& \frac{12 \epsilon}{\beta l_x^3 l_z}  \int_{-l_x/2}^{l_x/2} s_\psi \xt\,  (1 + f )  \, \cos{(x_c - t + c_\psi  \xt)}  \, h (\xt,t)  \,  d\xt . \label{psi_int}
\end{eqnarray}
\ese
This system of equations  \eqref{nondimprob} is sufficient to calculate the second order mean yaw moment. We find an asymptotic solution in the small wave limit $\epsilon \rightarrow 0$. This means that variables $x_c, \psi, f,h $ are expanded in powers of $\epsilon$:
\bse
\begin{eqnarray}
x_c&=& \overline{x}_c + \epsilon x_c' + \epsilon^2 x_c'' \\
\psi &=& \overline{\psi} + \epsilon \psi' + \epsilon^2 \psi'' \\
f & = & \overline{f} + \epsilon f ' + \epsilon^2 f '' \\
h & = & \overline{h}+ \epsilon h' + \epsilon^2 h'' 
\end{eqnarray}
\ese
As usual in multi-scale analysis, we also admit that these variables can vary on multiple time-scales, meaning that 
\be
\dot{\ } \rightarrow  \pd_t  + \epsilon \pd_\tau  +  \epsilon^2 \pd_T .
\ee
We inject these expansions in the equations and collect equations of motion at different orders of $\epsilon$. We limit the calculation of \eqref{xc_int} to order $\epsilon^1$, so we can simplify $\cos{(x_c - t + c_\psi  \xt)} \approx  \cos{(\ovl{x}_c - t + \ovl{c}_\psi  \xt)} $ in the integrand. The equation for the yaw angle   \eqref{psi_int} is wanted up to order  $\epsilon^2$, so there we need to use the Taylor expansions
\bse
\begin{eqnarray}
s_\psi &=& \ovl{s}_\psi + \epsilon \psi'  \ovl{c}_\psi + O (\epsilon^2) \\
\cos{(x_c - t + c_\psi  \xt)} &= & \cos{(\ovl{x}_c - t + \ovl{c}_\psi  \xt)} + \epsilon (- x_c'+ \psi'  \ovl{s}_\psi \wt{x})  \sin{(\ovl{x}_c - t + \ovl{c}_\psi  \xt)}  +  O (\epsilon^2 )   
\end{eqnarray}
\ese 
We denote $\ovl{s}_\psi = \sin \ovl{\psi}$ and $\ovl{c}_\psi = \cos \ovl{\psi}$. In the following, we first present the asymptotic solution for floaters with arbitrary lengths and then we study the short floater limit. 

\subsection{General theory for arbitrary lengths}

At order $\epsilon^0$, in absence of wave motion, the plate is flat and we also have $\pd^2_{tt} \ovl{x}_c = 0, \pd^2_{tt} \ovl{\psi} = 0$. The equilibrium is
\be
\ovl{x}_c, \ovl{\psi} \ \ \text{arbitrary} , \quad  \overline{f} = - \beta l_z  \ \ \text{and} \ \  \overline{h} = \beta l_z  .
\ee
At order $\epsilon^1$, the first order plate deformation $f'$ satisfies
\be \label{plate_eq}
l_D^4 \frac{\pd^4 f '}{\pd \wt{x}^4}  + f' =  \sin{(\ovl{x}_c - t +  \ovl{c}_\psi \xt )}  \ \ \text{with} \ \ \left . \frac{\pd^2 f'}{\pd \wt{x}^2} \right |_{\wt{x}= \pm l_x /2} =  \left . \frac{\pd^3 f '}{\pd \wt{x}^3} \right |_{\wt{x} = \pm l_x /2} = 0  .
\ee
The solution is 
\begin{eqnarray}
f ' &=& \left( \frac{\cos(\ovl{c}_\psi  \wt{x} )}{l_D^4 \ovl{c}_\psi^4 + 1} + A \cosh  \left ( \frac{j \wt{x}}{l_D}\right ) +  A^* \cosh  \left (  \frac{j^* \wt{x}}{l_D}\right ) \right ) \sin (\ovl{x}_c - t) \nonumber \\
& + &    \left( \frac{\sin(\ovl{c}_\psi \wt{x}  )}{l_D^4 \ovl{c}_\psi^4 + 1} + B \sinh  \left (  \frac{j \wt{x}}{l_D} \right )+  B^* \sinh  \left (  \frac{j^* \wt{x}}{l_D} \right )\right  ) \cos (\ovl{x}_c - t) ,  
\end{eqnarray}
where we denote $j= \exp (i \pi/4) = (1+ i )/\sqrt{2}$. The coefficients  $A, B$ are fixed by the boundary conditions, that lead to the linear systems
\bse \label{linsysAB}
\begin{equation}
\left [ \begin{array}{cc}
j^2 \cosh \left ( \frac{j l_x}{2 l_D} \right )  & {j^*}^2 \cosh \left ( \frac{j^* l_x}{2 l_D} \right )  \\
j^3 \sinh \left ( \frac{j l_x}{2 l_D} \right )  & {j^*}^3 \sinh \left ( \frac{j^* l_x}{2 l_D} \right ) 
\end{array} \right ]  \left [ \begin{array}{c}
A \\ A^* \end{array} \right ]  = \left [ \begin{array}{r}
\frac{l_D^2 \ovl{c}_\psi^2}{l_D^4 \ovl{c}_\psi^4 + 1} \cos \left (\frac{\ovl{c}_\psi l_x}{2} \right )   \\ - \frac{l_D^3 \ovl{c}_\psi^3}{l_D^4 \ovl{c}_\psi^4 + 1} \sin \left (\frac{\ovl{c}_\psi l_x}{2} \right )  \end{array} \right ] 
\end{equation}
and
\begin{equation}
\left [ \begin{array}{cc}
j^2 \sinh \left ( \frac{j l_x}{2 l_D} \right )  & {j^*}^2 \sinh \left ( \frac{j^* l_x}{2 l_D} \right )  \\
j^3 \cosh \left ( \frac{j l_x}{2 l_D} \right )  & {j^*}^3 \cosh \left ( \frac{j^* l_x}{2 l_D} \right ) 
\end{array} \right ]  \left [ \begin{array}{c}
B \\ B^* \end{array} \right ]  = \left [ \begin{array}{r}
\frac{l_D^2 \ovl{c}_\psi^2}{l_D^4 c_\psi^4 + 1} \sin \left (\frac{\ovl{c}_\psi l_x}{2} \right )   \\  \frac{l_D^3 \ovl{c}_\psi^3}{l_D^4 \ovl{c}_\psi^4 + 1} \cos \left (\frac{\ovl{c}_\psi l_x}{2} \right )  \end{array} \right ] .
\end{equation}
\ese
Explicit solutions for $A$ and $B$ are too complex to allow physical insight, but they can be easily calculated numerically for given, numerical values of $ \overline{\psi},l_x, l_D$. The first order deviation in submersion depth $h'$ is rewritten as  
\be \label{hprime}
h'  = \sin ({\xb}_c - t + \ovl{c}_\psi \xt ) - f ' = h_s' \sin  ({\xb}_c - t)  + h_c' \cos  ({\xb}_c - t)
\ee
and we have
\bse \label{hp}
\begin{eqnarray}
h_s'  &=&  \frac{ l_D^4 \ovl{c}_\psi^4}{l_D^4 \ovl{c}_\psi^4 + 1}  \cos (\ovl{c}_\psi \xt)  {-} 2 \text{Re} \left ( A \cosh  \left ( \frac{j \wt{x}}{l_D}\right )  \right ) \\
h_c'  &=&  \frac{ l_D^4 \ovl{c}_\psi^4}{l_D^4 \ovl{c}_\psi^4 + 1}  \sin (\ovl{c}_\psi \xt)  {-}2 \text{Re} \left ( B \sinh  \left ( \frac{j \wt{x}}{l_D}\right )  \right ) .
\end{eqnarray}
\ese
The first order motion of $x_c'$ and $\psi'$  is determined by
\bse
\begin{eqnarray}
\pd_{tt}^2 x'_c& \approx &-  \frac{1}{l_x } \int_{-l_x/2}^{l_x/2}  \cos{(\ovl{x}_c - t +  \ovl{c}_\psi \xt) } \, d\xt  \\
\pd_{tt}^2 \psi'  &=&  \frac{12}{ l_x^3}  \int_{-l_x/2}^{l_x/2} \ovl{s}_\psi \xt\, \left  (1+ \ovl{f} \right  ) \cos{(\xb_c - t +\ovl{c}_\psi  \xt )}  \, d\xt .
\end{eqnarray}
\ese
In the equation for $\psi'$, we can approximate $1 +\ovl{f} \approx 1$, because the floater is thin ($\ovl{f} = - \beta l_z$ and $\beta l_z \ll 1$). Both integrals can be analytically calculated and lead to the solutions
\be \label{firstordermotion}
  x'_c = \sinc \left ( \frac{\overline{c}_\psi l_x}{2}\right )  \cos(\xb_c - t), \quad \psi'  =  \frac{12}{l_x^2} \frac{\partial}{\partial \overline{\psi}} \left[  \sinc \left ( \frac{\overline{c}_\psi l_x}{2}\right ) \right] \sin(\xb_c - t),
\ee
with \text{sinc} the cardinal sinus function. It is important to notice that the first order motion of $x_c'$ and $\psi'$ is independent of the bending modulus: {\it flexible and rigid floaters have the same first order motion in horizontal position and yaw angle}. This is a consequence of the fact that the equilbrium submersion is independent of bending modulus. Ref.~\cite{dhote2025flexible} also found the same expressions for $x_c'$ and $\psi'$, in the perfectly flexible limit.  
\begin{figure}
 \centering
    \includegraphics[width = 0.9\linewidth]{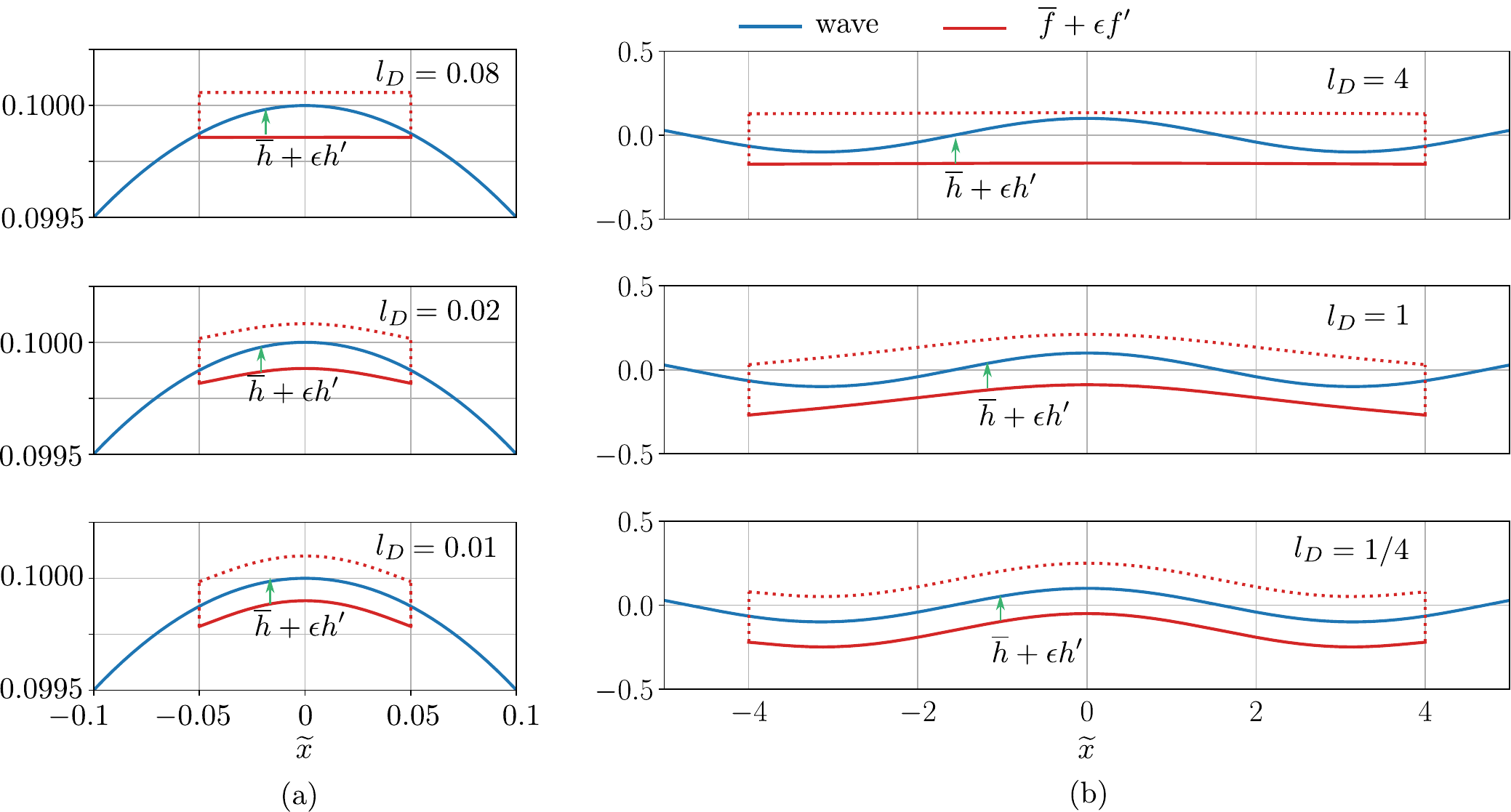}
    \caption{Deformation $  \overline{f} + \epsilon f' $ of short and long elastic floaters with $\beta=0.5$ and varying flexural lengths $l_D$ in a $\epsilon=0.1$ wave. (a) short floater $(l_x,l_z)=(0.1,2 \times 10^{-4})$, (b) long floater $(l_x,l_z)=(8,0.3)$. The phase is fixed to center the floater on a crest ($\sin ( \ovl{x}_c - t) =1$) and we aligned the long axis with the $x$-axis of wave-propagation  $(\ovl{\psi} = 0)$ to see maximal deformation. 
    \label{hsc}}
\end{figure}

In figure \ref{hsc}, we show a few examples of calculated non-dimensional deformations $\overline{f}+\epsilon f'$ for a short (a) and a long (b) elastic floater and for varying flexural length $l_D$. We fix $\epsilon=0.1$ and with $\beta = 0.5$ the equilibrium submersion depth is $l_z/2$. We fix the phase so that $\sin( \ovl{x}_c-t) = 1$ and align the floater with the $x$-axis ($\ovl{\psi}=0$).
For flexural lengths that are of the order of strip length or greater, the elastic strip is very stiff and remains nearly undeformed. The spatial variation of submersion  $h'$  along the long axis is then very significant. For flexural lengths that are significantly smaller than the strip length, we see that the elastic strip adapts to the surface, keeping a nearly constant submersion along the long axis, which implies $h' \rightarrow 0$.

We proceed our calculation and write the order $\epsilon^2$ problem for the yaw angle. Simplifying $1 +\ovl{f} \approx 1$ as before, we find
\begin{eqnarray}
\pd^2_{\tau\tau} \ovl{\psi} + \pd^2_{tt}  {\psi}''  &=& \frac{12}{l_x^3}  \int_{-l_x/2}^{l_x/2} x_c'  \left (- \ovl{s}_\psi   \xt  \sin{(\ovl{x}_c - t + \ovl{c}_\psi  \xt)}  \right ) \,  d\xt  \nonumber \\
& +& \frac{12}{l_x^3}  \int_{-l_x/2}^{l_x/2} \psi'  \left (   \ovl{c}_\psi   \xt \cos{(\ovl{x}_c - t + \ovl{c}_\psi  \xt)}  +  \ovl{s}_\psi^2   \wt{x}^2 \sin{(\ovl{x}_c - t + \ovl{c}_\psi  \xt)}  \right  )  \, d\xt \nonumber \\
& + & \frac{12}{\beta l_x^3l_z}  \int_{-l_x/2}^{l_x/2}   h'   \left ( \ovl{s}_\psi  \xt    \cos{(\ovl{x}_c - t + \ovl{c}_\psi  \xt)}  \right ) \,  d\xt \nonumber \\
& +&  \underbrace{ \frac{12}{l_x^3}  \int_{-l_x/2}^{l_x/2}  f '  \left ( \ovl{s}_\psi  \wt{x} \cos{(\ovl{x}_c - t + \ovl{c}_\psi  \xt)}  \right )   \,d\xt .}_{\text{negligible}}  \label{eqpsilent}
\end{eqnarray}
We inject the first order solutions  $x_c',  \psi',h', f '$ in the right hand side and average over the short time-scale (denoted using overline). The fourth term is negligible, because its average is exactly $-\beta l_z$ times the average of the third term. 
We introduce the notation  
\bse
\begin{eqnarray}
\pd^2_{\tau\tau} \ovl{\psi}  &=& \underbrace{\ovl{\mathcal{K}}_z^{L} \left (\ovl{\psi},l_x \right ) + \ovl{\mathcal{K}}_z^{T} \left (\ovl{\psi},l_x,\beta l_z, l_D \right )}_{\overline{\mathcal{K}}_z}  \label{psieq}
\end{eqnarray}
that splits the non-dimensional mean yaw moment $\ovl{\mathcal{K}}_z$ in two parts, labeled (L) and (T),  
\ba
\ovl{\mathcal{K}}_z^{L}   \left (\ovl{\psi},l_x \right )  &=& \frac{12}{l_x^3}  \int_{-l_x/2}^{l_x/2} \overline{x_c'  \left (- \ovl{s}_\psi   \xt  \sin{(\ovl{x}_c - t + \ovl{c}_\psi  \xt)}  \right )} \,  d\xt  \nonumber \\
& + &  \frac{12}{l_x^3}  \int_{-l_x/2}^{l_x/2}  \overline{\psi'  \left (   \ovl{c}_\psi   \xt \cos{(\ovl{x}_c - t + \ovl{c}_\psi  \xt)}  +  \ovl{s}_\psi^2   \wt{x}^2 \sin{(\ovl{x}_c - t + \ovl{c}_\psi  \xt)}  \right  )}  \, d\xt  \label{KzLdef}
\ea
and 
\ba
\ovl{\mathcal{K}}_z^{T}   \left (\ovl{\psi},l_x \right )  &=& \frac{12}{\beta l_x^3l_z}  \int_{-l_x/2}^{l_x/2}   \overline{h'   \left ( \ovl{s}_\psi  \xt    \cos{(\ovl{x}_c - t + \ovl{c}_\psi  \xt)}  \right )} \,  d\xt . \label{KzTdef}
\ea
\ese
Indices L and T refer to the fact that they respectively favor the longitudinal or the transverse position in the short floater limit.  

Using the first order motion $x_c'$ and $\psi'$ in these integrals, we can calculate the L-part analytically and express it in terms of $\text{sinc}$-functions or with spherical Bessel functions $j_n (X)$: 
\begin{eqnarray}
\ovl{\mathcal{K}}_z^{L}   \left (\ovl{\psi},l_x \right )  &=&  - \frac{6}{l_x^2} \frac{\pd}{\pd \ovl{\psi}} \left ( \sincxb \right ) \left [  \sincxb   + \frac{12}{l_x^2} \frac{\pd^2}{\pd \ovl{\psi}^2} \left ( \sincxb \right )    \right ] \nonumber  \\
& =&   -    \frac{6 \ovl{s}_\psi }{l_x} j_1 \left ( \frac{\ovl{c}_\psi l_x}{2}\right ) \left [ \ovl{c}_\psi^2  j_0  \left ( \frac{\ovl{c}_\psi l_x}{2} \right ) + \left (1 - \frac{\ovl{c}_\psi^2}{2}  \right ) 
j_2 \left ( \frac{\ovl{c}_\psi l_x}{2}\right ) \right ]\label{Kzlong}
\end{eqnarray}
We used here the relations $\text{sinc} ( X) = j_0 (X)$, $\text{sinc}' ( X) =- j_1 (X)$,  $ 3\,  \text{sinc}'' ( X) = 2  j_2 (X) -  j_0 (X)$. This L-part of the non-dimensional mean yaw moment only varies with angle of incidence $\ovl{\psi}$ and non-dimensional floater length $l_x$. It is also  identical to the mean yaw moment found in \cite{dhote2025flexible} for perfectly flexible floaters. 

The T-part is due to the spatially varying submersion that varies with bending rigidity. We can express the integral as 
\begin{eqnarray}\label{Kztrans}
\ovl{\mathcal{K}}_z^{T}  \left (\ovl{\psi},l_x, \beta l_z , l_D \right )&=  &\frac{6}{\beta l_x^3 l_z}  \int_{-l_x/2}^{l_x/2}    \left ( h_c'  \ovl{s}_\psi  \xt  \cos{( \ovl{c}_\psi  \xt)} - h_s'  \ovl{s}_\psi  \xt \sin{( \ovl{c}_\psi  \xt)}  \right )  \,  d\xt  \end{eqnarray}
and we need to inject here, the profiles $h_c'$ and $h_s'$ identified in \eqref{hp}. These functions have a non-trivial dependence on the flexural length $l_D$ and so we evaluate this integral numerically using a simple quadrature rule (see the Jupyter notebook in the Supplementary Material).

\begin{figure}
 \centering
    \includegraphics[width = 0.8\linewidth]{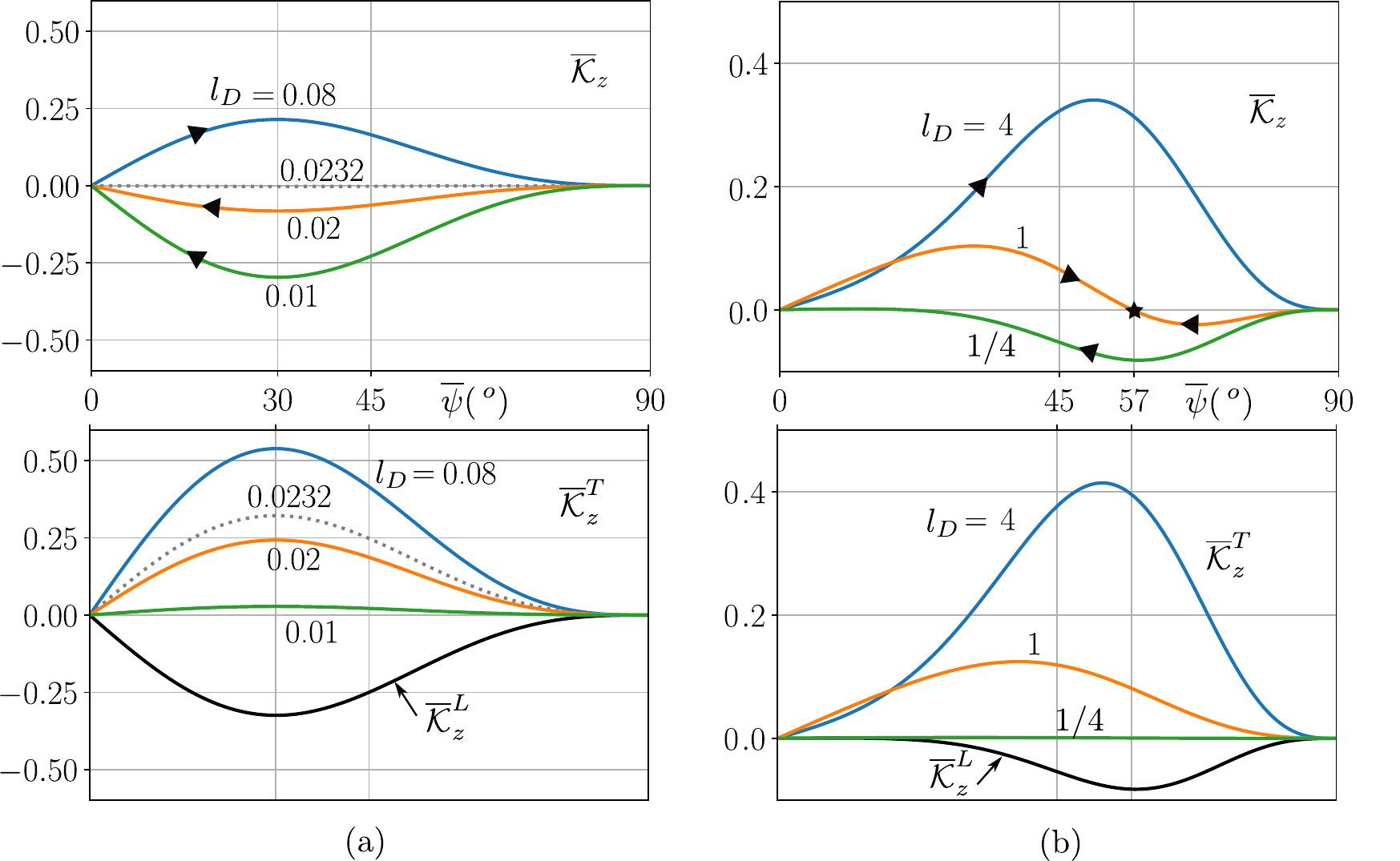}
    \caption{Non-dimensional mean yaw moment $\overline{\mathcal{K}}_z$ and competing parts $\overline{\mathcal{K}}_z^{L}$, $\overline{\mathcal{K}}_z^{T}$, as a function of angle of incidence $\ovl{\psi}$ for the floaters of figure \ref{hsc} with $\beta=0.5$ and varying flexural length $l_D$.  (a) Short elastic floater $(l_x , l_z) = (0.1, 2 \times 10^{-4})$, showing two equilibrium angles, $\ovl{\psi} = 0$ or $90^\circ$;  (b) long elastic floater  $(l_x , l_z) = (8, 0.3)$, with an additional equilibrium angle (here at $57^\circ$). Blacks arrows indicate the slow motion of $\ovl{\psi}$ towards the stable equilibria.
    \label{Kz_visu}}
\end{figure}
As a first numerical application, we have calculated the non-dimensional mean yaw moment  $ \ovl{\mathcal{K}}_z$ and the separate L and T-parts, as a function of angle of incidence $\ovl{\psi} \in [0, 90^o]$, for the same floaters as those of figure \ref{hsc}.  

In figure  \ref{Kz_visu}(a), we consider the case of the short floater with $l_x=0.1$. We see that $\ovl{\mathcal{K}}_z = 0$ in both the longitudinal $\ovl{\psi} = 0^o$ and the transverse position $\ovl{\psi} = 90^o$ and that there are no other equilibria. We also see that the mean yaw moment is either positive or negative over the entire $\ovl{\psi}$ interval, with a sign that depends on $l_D$. The black arrows in this diagram suggest the direction in which angular drift will occur, so they point towards the stable, preferential orientation. When $\ovl{\mathcal{K}}_z <0$, the floater will slowly rotate to the longitudinal state $\ovl{\psi}=0^o$ (for the more flexible floaters, $l_D =0.01, 0.02$). When $\ovl{\mathcal{K}}_z >0$, the floater will instead rotate towards the transverse state $\ovl{\psi}=90^o$ (for the more rigid floater, $l_D=0.08$). Bending rigidity clearly influences preferential orientation. In this numerical application, the mean yaw moment changes sign near $l_D=0.023$ and this precisely locates a transition in preferential orientation. In the bottom row, we see that the T-part of the moment vanishes for small $l_D$, as expected in the very flexible limit: submersion is no longer varying, $h' \rightarrow 0$. We can finally notice that all curves have similar shapes and this is normal: below we show that $\ovl{\mathcal{K}}_z \sim \sin \ovl{\psi} \cos^3 \ovl{\psi}$ in the small floater limit. 

In figure  \ref{Kz_visu}(b), we consider the longer elastic floater with $l_x=8$. The bottom row suggests that we still have a T-part that is mostly positive and a L-part that is mostly negative. However, we also see changing shapes in the curves of the T-part as we vary $l_D$. The result is a total yaw moment $\ovl{\mathcal{K}}_z$ that has a more complex $\overline{\psi}$-dependence. For the flexural length $l_D=1$, we can see that a new stable equilibrium orientation ($\overline{K}_z=0$) emerges near $\ovl{\psi}=57^o$ ($\star$ in figure). Both the longitudinal and the transverse equilibrium are unstable for this parameter set. This is an example that shows that longer floaters can also stabilise at intermediate angles.

\begin{figure}[h!]
 \centering
    \includegraphics[width = 0.8\linewidth]{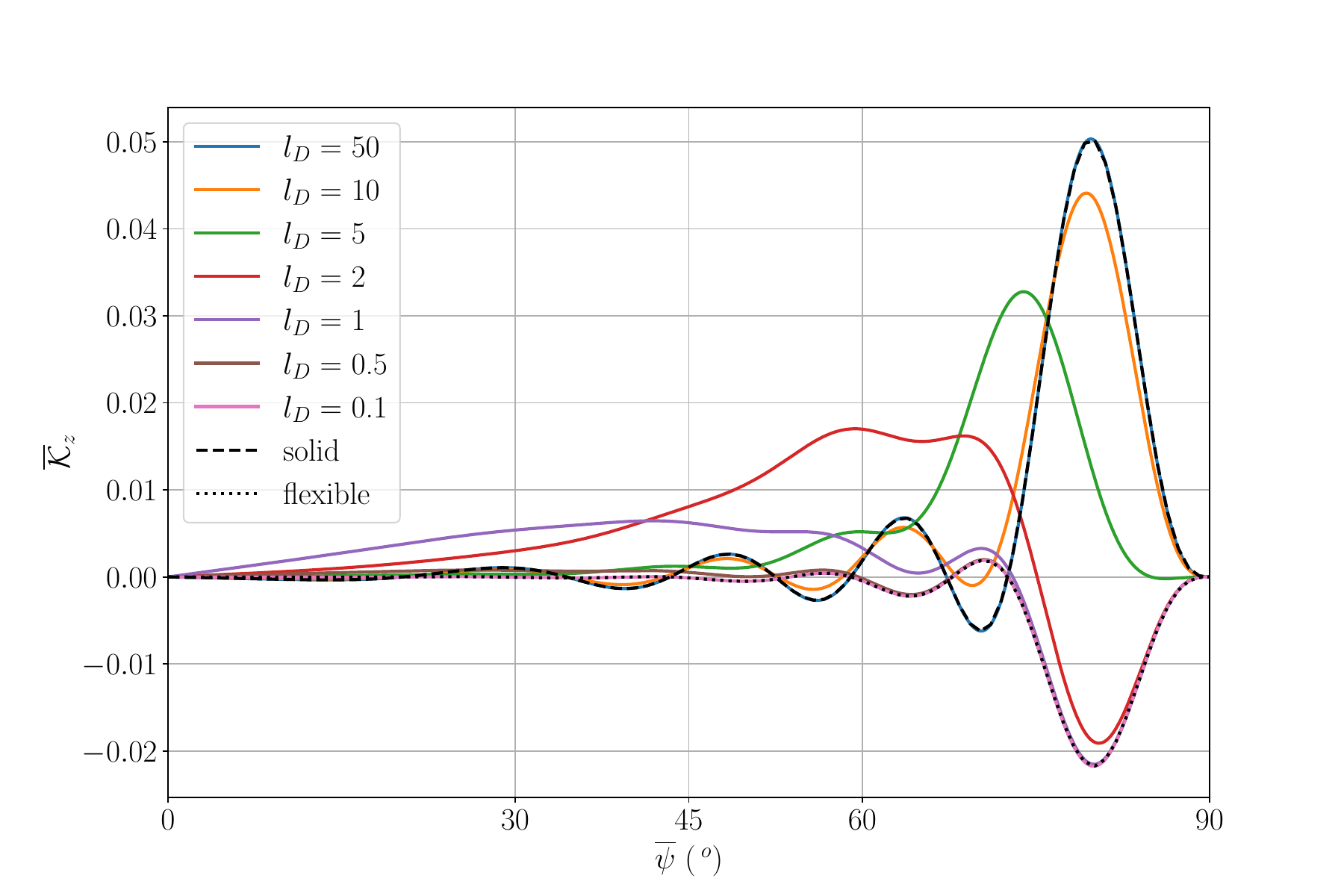}
    \caption{The non-dimensional mean yaw moment $\overline{\mathcal{K}}_z$ on very long floaters ($l_x=30$, $l_z=0.6$ and $\beta=0.5$) shows complex variation with angle $\ovl{\psi}$ and flexural length $l_D$, but we recover the expected asymptotic behavior in the perfectly flexible and solid limits. There can be multiple stable positions where $\overline{\mathcal{K}}_z=0$ and preferential orientation may depend on initial conditions or experimental noise.  } 
    \label{Kz_verylong}
\end{figure}
In Figure \ref{Kz_verylong}, we consider another numerical application in which we calculate the mean yaw moment as a function of angle for a very long floater with  $l_x = 30$, nearly 5 wavelengths long. We vary $l_D$ in a broad range. We also add the flexible limit in which the T-part of mean yaw moment is zero and the solid limit formula that is discussed in Appendix \ref{sec:appendix}. The main message of this figure is that the variation of the mean yaw moment with angle and $l_D$ becomes very complex in the case of very long floaters. In most curves, we can see that there are multiple angles of equilibrium where $\ovl{K}_z=0$ and almost everywhere, the non-dimensional mean yaw moment is significantly reduced in magnitude. As a result, the preferential state of orientation of very long floaters can be very case-specific and also dependent on initial conditions. Because of this it is no longer very meaningful to speak of preferential orientation in the case of very long floaters. The resulting orientation may well be unpredictable. Physically, the complex variation of the mean yaw moment for very long floaters is the result of compensating pressure forces. Along its axis, the floater is feeling a dynamical pressure wave with a projected wavelength $\lambda / \cos \ovl{\psi}$. When  $\lambda / \cos \ovl{\psi} > L_x$, the wave is pushing both back and forward on the surface of floater and this causes opposing contributions in the non-dimensional mean yaw moment (and mean drift force). This explains why there is a significant reduction in magnitude of $\overline{\mathcal{K}}_z$ and also why multiple angles of equilibrium can exist. 

\subsection{Short floater or long wavelength limit}

In the short-floater  or long wavelength limit $ \lambda \gg L_x \gg L_y \gg L_z$, there is a clear preference for either the longitudinal or transverse state of orientation. We simplify the calculation of the mean yaw moment in this limit and find the criterion \eqref{prediction}. 

Using the small argument $X \ll 1$ expansions of the spherical Bessel functions, $ j_0(X) \approx 1$, $j_1 (X) \approx X/3 $ and $ j_2 (X) \approx X^2/15$ in the L-part of the mean yaw moment \eqref{Kzlong}, we directly obtain
\be \label{Kzsmall_part1}
\overline{\mathcal{K}}_z^{L} \approx - \ovl{s}_\psi \ovl{c}_\psi^3.
\ee
To simplify the T-part of the mean yaw moment, we reconsider the calculation of the strip deformation $f '$. In the right hand side of equation \eqref{plate_eq}, we replace $\sin{(\ovl{x}_c - t +  \ovl{c}_\psi \xt )} $ with a second order Taylor expansion along $\wt{x}$:
\be \label{deform_small}
l_D^4 \frac{\pd^4 f '}{\pd \wt{x}^4}  + f' = \left (1 - \frac{\ovl{c}_\psi^2 \xt^2}{2} \right ) \sin (\ovl{x}_c-t ) +   \overline{c}_\psi \xt \cos (\ovl{x}_c-t ) .
 \ee
In the short floater limit, dynamical pressure is always enforcing a parabolic variation on the short elastic floater. The solution for the first order deformation is  
\begin{eqnarray}
f ' &=& \left (1 - \frac{\ovl{c}_\psi^2 \xt^2}{2}    + l_D^2 \ovl{c}_\psi^2 \left (  \mathcal{A}  \cosh  \left ( \frac{j \wt{x}}{l_D}\right ) +   \mathcal{A}^* \cosh  \left (  \frac{j^* \wt{x}}{l_D}\right ) \right ) \right ) \sin (\ovl{x}_c - t) +  ( \overline{c}_\psi \xt ) \cos (\ovl{x}_c-t ) \nonumber 
\end{eqnarray}
with 
\be
\mathcal{A} = \frac{ \sqrt{2} j^* \sinh \left(  \dfrac{j^*l_x}{2l_D} \right )}{\sinh \left ( \dfrac{l_x}{\sqrt{2} l_D}  \right ) + \sin \left ( \dfrac{l_x}{\sqrt{2} l_D}  \right ) }.
\ee
The first order deviation in local submersion depth $h'$ is as in Eq.~\eqref{hprime}, but we now have 
\be
  h_s' = -   l_D^2 \ovl{c}_\psi^2 \, 2 \text{Re} \left (  \mathcal{A}  \cosh  \left ( \frac{j \wt{x}}{l_D}\right ) \right)  \ , \ h_c' = 0.
\ee
Injecting these profiles in the integral \eqref{Kztrans} and replacing  $\sin{( \ovl{c}_\psi  \xt)} \approx  \ovl{c}_\psi  \xt$, we can calculate the integral analytically and find 
\ba
 \overline{\mathcal{K}}_z^{T}
&\approx & 
\ovl{s}_\psi \ovl{c}_\psi^3  \left [ \frac{12 \, l_D^4}{\beta l_z l_x^2} - \frac{24 \sqrt{2} \, l_D^5}{\beta l_z l_x^3} \frac{\cosh \left (   \dfrac{l_x}{\sqrt{2} l_D}      \right )   }{\sinh \left (   \dfrac{l_x}{\sqrt{2} l_D}   \right ) +\sin \left (   \dfrac{l_x}{\sqrt{2} l_D}   \right ) } \right ]. \label{Kzsmall_part2}
  \ea
In the short limit, both L and T parts of the mean yaw moment vary with angle as $\sin \ovl{\psi} \cos^3 \ovl{\psi}$.  Combining \eqref{Kzsmall_part1} and \eqref{Kzsmall_part2} we find the simplified equation of motion for the mean yaw angle in the small floater limit:
 \bse
\begin{eqnarray}\label{crit_short}
 \pd^2_{\tau\tau} \ovl{\psi}  &=& \ovl{s}_\psi  \ovl{c}_\psi^3 \left ( -1  + \frac{F}{F_c}  \right ) \label{eqpsismall}
\end{eqnarray}
with
\be
F = \frac{l_x^2}{\beta l_z}  = \frac{k L_x^2}{\beta L_z}
\ee
and a critical value $F_c$ that depends on the ratio $l_D/l_x= L_D/L_x$:
\be \label{Fcdef}
F_c  = \left [  3 \left (  \frac{\sqrt{2} L_D}{L_x} \right )^4  - 6  \left (  \frac{\sqrt{2} L_D}{L_x} \right )^5 \, \, \frac{\cosh \left (   \dfrac{L_x}{\sqrt{2} L_D}   \right ) - \cos \left (   \dfrac{L_x}{\sqrt{2} L_D}   \right )   }{\sinh \left (   \dfrac{L_x}{\sqrt{2} L_D}   \right ) + \sin \left (   \dfrac{L_x}{\sqrt{2} L_D}   \right )   }   \right ]^{-1}.
\ee
\ese
This result \eqref{crit_short} is similar to that of Refs.~\cite{herreman2024preferential,dhote2025flexible} and allows the simple interpretation given in the introduction. The sign of the non-dimensional mean yaw moment in the right hand side of \eqref{eqpsismall} is controlled by the non-dimensional number $F$ and how it compares to a critical value $F_c$. For $F < F_c$, the mean yaw moment is negative and the longitudinal position ($\ovl{\psi}= 0^o$) is stable. For $F > F_c$, the mean yaw moment is positive and the transverse position ($\ovl{\psi}= 90^o$) is stable. Using the small and large argument expansions of the trigonometric and hyperbolic functions, we can get the asymptotic limits
\bse
\begin{eqnarray}
  L_x/L_D \rightarrow 0 & :&  F_c  \approx  60 + {\frac{5}{42}}  \left (  \frac{L_x}{L_D} \right )^{4} \label{Fc_low}\\ 
  L_x/L_D \rightarrow \infty & :&  F_c \approx  \frac{1}{12} \left (  \frac{L_x}{L_D} \right )^{4} .  \label{Fc_high}  
\end{eqnarray}
\ese
For rigid floaters ($L_x/L_D \ll 1$) we find the limit $F_c \rightarrow 60$. For very flexible floaters ($L_x/L_D\gg 1$), $F_c$ grows as $(L_x/L_D)^4  = \rho g L_x^4 / D$, inversely proportional to bending rigidity. Note that this scaling law is found in both low and high $L_x/L_D$ asymptotic expansions of  equations \eqref{Fc_low} and \eqref{Fc_high}, but with a slightly different prefactor ({$5/42 \simeq 0.119$, versus $1/12 \simeq 0.083$}).
\begin{figure}[tb]
 \centering
    \includegraphics[width = 0.8\linewidth]{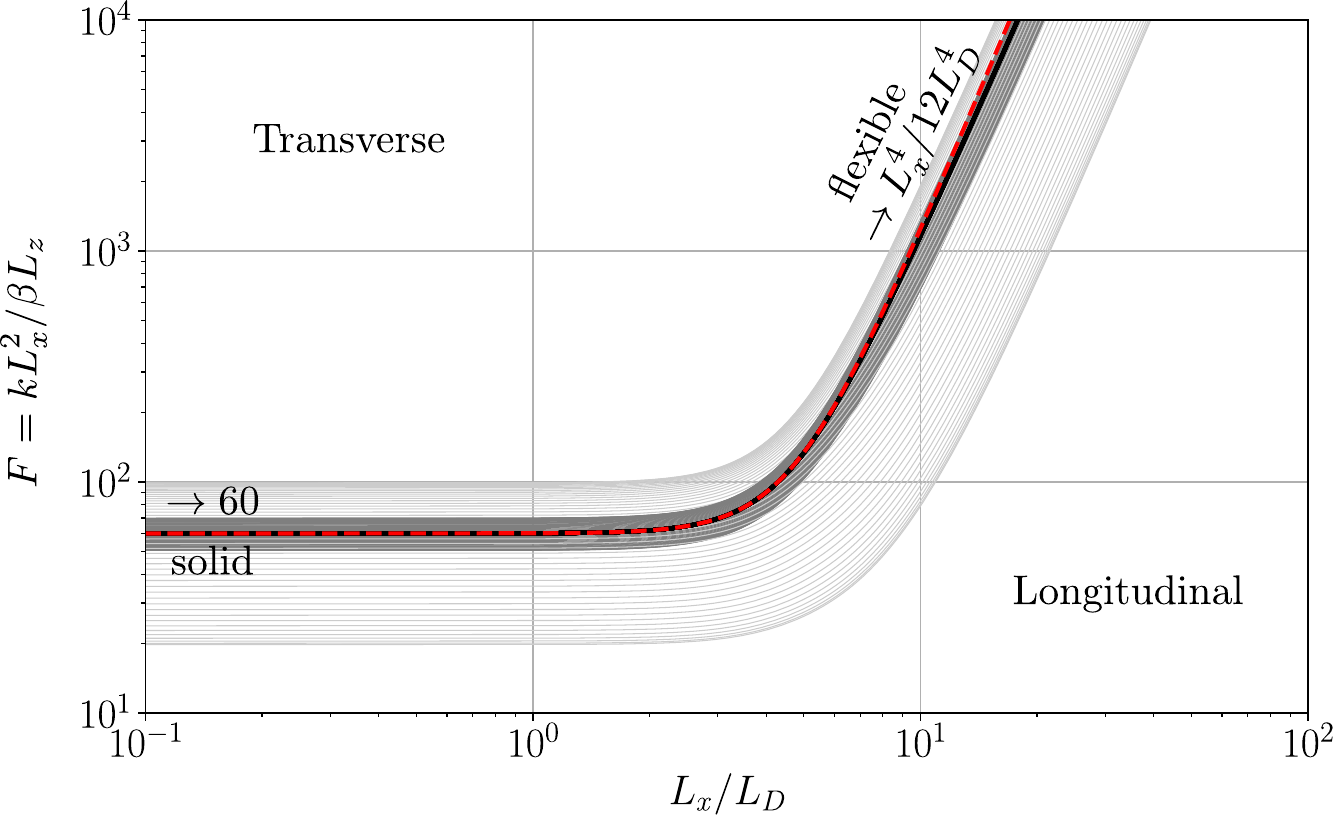}
    \caption{Theoretical phase diagram for preferential orientation according to the short floater theory. The transition line $F_c$ (Eq.~\ref{Fcdef})  is shown as the black line. For $F<F_c$, longitudinal orientation is stable, whereas for $F>F_c$ transverse orientation is stable.  The critical value $F_c$ asymptotes towards $60$ for rigid floaters, and to $L_x^4/12 L_D^4$ for elastic floaters. The red dashed line shows the small $L_x/L_D$ expansion (\ref{Fc_low}), that almost superimposes to the arbitrary-length theory. Dark and light gray correspond to uncertainty ranges around the transition line  or intermediate equilibrium positions,  that appear as we move beyond the short floater limit ($L_x/\lambda = 1/2$ dark gray, $L_x/\lambda = 1$ light gray). The short floater theory applies when $L_x/\lambda  < 1/2$.  }
    \label{FIG_transit_short}
\end{figure}
In figure \ref{FIG_transit_short}, we show the critical $F_c$ (\ref{Fcdef}) in the  ($L_x/L_D, F$) plane as a black line. When $F < F_c$, we expect longitudinal orientation, whereas for $F> F_c$ we expect a transverse orientation. The red dotted line is the simpler approximation  \eqref{Fc_low} and it provides an accurate and convenient description of the transition over the entire range of the diagram.

In the short-floater theory, the wave profile is approximated by a parabola, which remains appropriate for lengths up to $L_x \leq \lambda/2$. Beyond half a wavelength, it is necessary to account for the sinusoidal shape of the wave. This causes the more complex variation of the mean yaw moment with angle of incidence observed in figures \ref{Kz_visu}(b) and \ref{Kz_verylong}, thereby rendering the prediction of a longitudinal versus transverse preferential orientation uncertain. We estimate how uncertainty appears as we leave the short limit using the following procedure. We compute, for a given choice of $l_x$ and $l_D$, the critical value of $F$ at which the mean yaw moment vanishes for each angle in the interval $\overline{\psi} \in [0,90^o]$. In the short limit, the mean yaw moment changes sign for all angles $\overline{\psi}$ at the same $F=F_c$, but for longer floaters, there is a different critical value $F_c$ for each angle due to the existence of intermediate equilibra. This defines a set of transition lines, which we plot in figure \ref{FIG_transit_short},  in dark gray for $l_x = \pi$ ($L_x = \lambda/2$) and light gray for $l_x = 2\pi$ ($L_x = \lambda$). Taken together, these lines define a range of uncertainty around the short-floater limit prediction  (black line). This diagram {confirms} that the short-floater theory  provides a good prediction for the preferential orientation up to $L_x \simeq \lambda/2$. Beyond this limit, we must use the general theory and complexer situations will arise. 

\begin{figure}
    \centering
    \includegraphics[width = 1\linewidth]{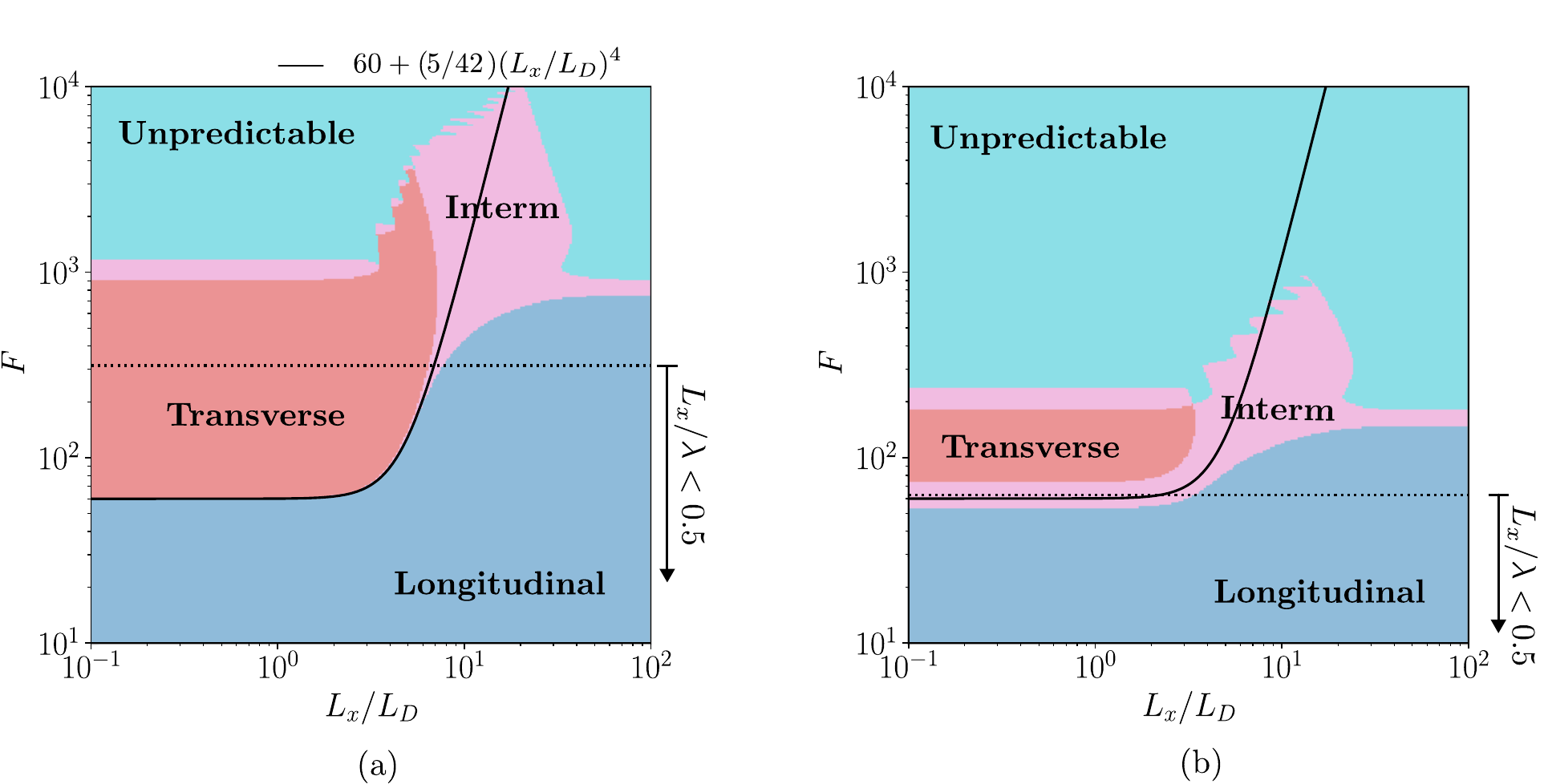}
    \caption{Phase diagrams in the $F$ and $L_x/L_D$ plan, based on the arbitrary length theory and for fixed ratios of length to submersion depth  $L_x / \ovl{h} = 100$ (a) and $20$ (b).  Patches distinguish regions of space where the longitudinal, the transverse or an intermediate equilibrium is stable. When more than 2 stable equilibria exist, we use the label unpredictable. The full black line gives the short limit prediction of the transition line \eqref{Fc_low} and correctly describes the longitudinal-transverse transition under the dotted line, for $L_x/\lambda<0.5$.   }
    \label{lxsld_fixe_diag}
\end{figure}

Predictions for the preferential orientation within the arbitrary-length theory are difficult to represent, as they now depend on three dimensionless parameters: $L_x/L_D$, $L_x/\overline{h}$, and $kL_x$ (with $F$ being the product of the latter two). The initial angle even enters as a fourth control parameter when one or more intermediate equilibrium angles exist. For practical applications, it is relevant to fix the ratio $L_x / \ovl{h}$ and examine the preferential orientation for varying $L_x/L_D$ and $F$, which is then directly proportional to $kL_x$. The corresponding phase diagrams are shown in figure \ref{lxsld_fixe_diag} for $L_x / \ovl{h}= 100$ and 20, values representative of typical floating structures. For each point in these diagrams, we determine the stable angles of equilibria, for which $\ovl{\mathcal{K}}_z = 0 $ and $\partial \ovl{\mathcal{K}}_z / \partial \ovl{\psi}  <0 $. Four regions are identified: (1) stable longitudinal equilibrium; (2) stable transverse equilibrium; (3) single intermediate equilibrium; (4) two or more intermediate equilibra. Region (4), that corresponds to the case of multiple zeros in $\ovl{\mathcal{K}}_z$ (as in figure \ref{Kz_verylong}), is labelled as unpredictable, as the preferential orientation can strongly depend on initial conditions in that case. The short-floater predicted transition line  is also shown (black line), and the dotted line indicates the part of parameter space where $L_x/\lambda <0.5$, where the short limit theory applies. 

The case $L_x /  \ovl{h} = 100$,  relevant for light floating structures such as inflatables, surfboards, pontoons, as well as some large scale floating flexible structures \cite{zhang_2022}, is illustrated in  figure \ref{lxsld_fixe_diag}(a). In this case, a clear prediction on the preferential orientation can be made over a significant portion of the diagram. Above the dotted line, the short-limit approximation no longer applies, and the boundary between the transverse and longitudinal regions opens up into a (pink) region where one intermediate equilibrium is stable, as illustrated in figure~\ref{Kz_visu}(b). For very long floaters such that $F > 1000$ (i.e., for $kL_x > 10$), no prediction can be made on the preferential orientation.

The case $L_x / \ovl{h} = 20$, more relevant to boats or heavy structures such as storage offshore structures, is illustrated in figure \ref{lxsld_fixe_diag}(b).  Here, the part of diagram where the short limit theory applies becomes very narrow (under the dotted line, below $F< 20 \pi $). In this region, the preferential orientation is systematically longitudinal; a small island of transverse preferential orientation survives close to the short-limit boundary, but in most of the diagram, we cannot make meaningful predictions on the preferential orientation.

\section{Applications:  pontoons, inflatables and foam mats}
\label{sec:applications}

\begin{table}[t]
\begin{ruledtabular}
\begin{tabular}{lccccc}
 & $L_x $ & $L_y$ & $\overline{h}$ & $L_D$ & $\lambda$\\
\hline
modular pontoons & 1 - 100 & 1 - 3 & 0.2 & 10 & 1 - 100 \\
inflatable paddleboards and kayaks   & 3 & 0.5 & 0.05 & 0.5 - 10 & 0.5 - 20 \\
foam mats  & 0.2 - 1  & 0.1 & 0.01  & 0.1 - 5 & 0.5 - 3 
 \end{tabular}
\end{ruledtabular}
\caption{Typical numerical values of lengths, widths, drafts, flexural lengths and wavelengths in units (m), used in numerical applications to pontoons, inflatable structures and polyethylene foam mats.}
\label{Table_num} 
\end{table}

We now apply our theory to three specific types of floating structures: flexible pontoons, drifting inflatables such as kayaks or paddle boards, and XPE (cross-linked polyethylene) foam mats that could be used for experimental validation of the theory. Table \ref{Table_num} gathers typical numerical values for length, width, draft, flexural length and wavelength.  

\subsection{Mean yaw moments on moored flexible pontoons}

\begin{figure}[t]
    \centering
  \includegraphics[width = .9\linewidth]{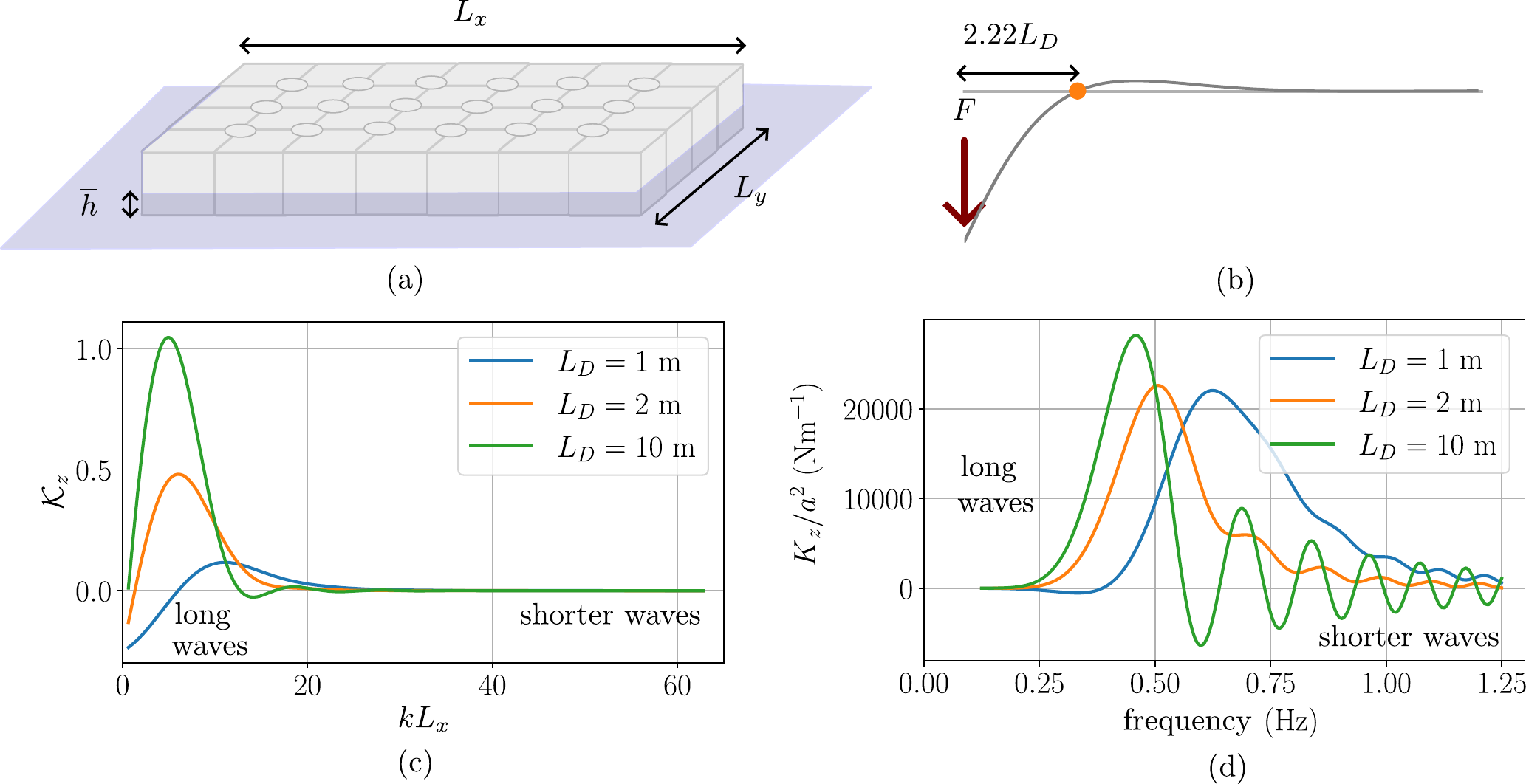}
    \caption{Mean yaw moments on flexible floating pontoons. (a) Sketch of a typical modular pontoon. (b) Determination of the flexural length by forcing downwards at one end of a sufficiently long pontoon.  (c) Non-dimensional mean yaw moment $\overline{\mathcal{K}}_z$  and (d)  dimensional mean yaw moment $\overline{K}_z/a^2$ relative to wave-amplitude squared as  function of $kL_x $ or wave-frequency.  Dimensions $(L_x,L_y, \overline{h} )=(10,1,0.1) $m, wavelength $\lambda \in [1, 100]$ m and varying flexural length $L_D=1,2, 10$~m.    \label{pontoon}}
    \end{figure}
Modular pontoons as in the sketch of figure \ref{pontoon}(a) are used in harbors, as temporary bridges or for security around wakeboard cable parks. They can be up to several hundreds of meters long and generally they are only a few meters wide, with a draft of a few tens of centimeters. The bending rigidity depends  on the way the individual modules are interconnected. We estimate $L_D$ to range from approximately $0.5$~m for highly compliant pontoons to about $20$~m for stiffer configurations. In practice, $L_D$ can be determined by pushing the pontoon downwards on one end and measuring the point of zero elevation, as illustrated in figure \ref{pontoon}(b). If thin plate theory applies, the resulting deformation is proportional to $\exp (-x / \sqrt{2} L_D) ) \cos (x / \sqrt{2} L_D)$, with the point of zero elevation (orange dot)  located at a distance $\pi/\sqrt{2}=2.22 L_D$ from the loaded end. 

Pontoons are rarely freely drifting so it is not meaningful to discuss preferential orientation. However, we can still use our theory to calculate typical mean yaw moments. An example is shown in figures  \ref{pontoon}(c) and (d),  for a length $L_x=10$ m, width $L_y=1$ m and draft $\ovl{h}=0.1$~m. We vary the wavelength in the interval $\lambda \in [1,100]$ m and fix the angle of incidence to $\ovl{\psi} =45^o$. In figure \ref{pontoon}(c), we show the non-dimensional mean yaw moment $\overline{\mathcal{K}}_z$ as a function of $kL_x$. For small $kL_x$ (long waves), $\overline{\mathcal{K}}_z$ is largest and its sign changes as a function of $L_D$, while for larger $kL_x$ (shorter waves), it rapidly decay. This non-dimensional mean yaw moment is a useful representation in  the preferential orientation problem as its magnitude measures the angular drift acceleration. However, for moored systems, the raw (not normalized by the moment of inertia) mean yaw moment is a more relevant quantity. In figure \ref{pontoon}(d), we show this dimensional mean yaw moment $\overline{{K}}_z/a^2$ normalized by the square of the incoming wave amplitude as a function of wave frequency, a more common representation in naval engineering applications. We see that the dimensional mean yaw moment remains largest for long waves, but its decay with frequency is less pronounced. Such data provide a useful benchmark for future hydro-elastic numerical simulations.

\subsection{Preferential orientation of drifting, inflatable structures in waves}

\begin{figure}[t]
    \centering
  \includegraphics[width = .9\linewidth]{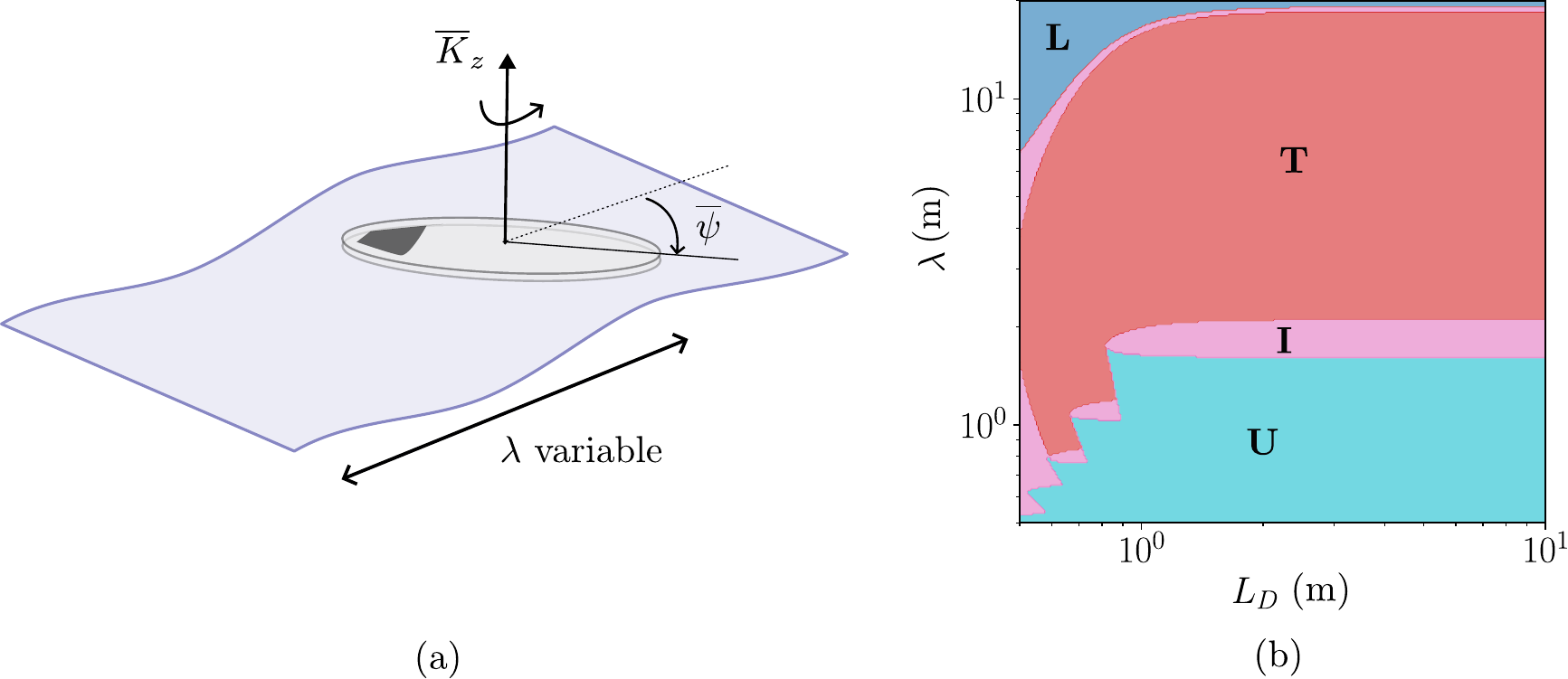}
    \caption{(a) Sketch of paddle board in waves. We vary wavelength and flexural length. (b) Phase diagram for preferential orientation (L = longitudinal, T = transverse, I = intermediate, U = unpredictable). 
        \label{application}}
    \end{figure}
    
Inflatable floating structures such as paddle boards, kayaks or floating mats are very common in recreational activities. They typically measure a few meters in length, less than a meter in width, and have a draft of at most a few centimeters. Such inflatable structures are generally flexible, although stiff configurations can be realized using technologies such as drop-stitch construction, which permits high-pressure inflation. The flexural length typically varies from a few tens of centimeters to a few meters. Placed in waves, as sketched in  figure \ref{application}(a), these inflatable structures are subjected to a mean yaw moment $\overline{K}_z$ that will act on their course. Since the typical velocity of a paddle board or kayak is usually smaller than the wave propagation velocity, the situation can be approximated as that of a freely drifting object. 

In figure \ref{application}(b), we show a phase-diagram for the preferential orientation of a structure with dimensions $(L_x,L_y,\overline{h}) = (3,0.5,0.05)$ m in waves with varying wavelength $\lambda \in [0.5,20] $ m and flexural lengths $L_D \in [0.5, 10]$ m.  Only for very flexible structures and in long waves, we predict a longitudinal preferential state. For  wavelengths in the range $\lambda \in [2,20]$ m, the transverse orientation is preferred. This preference for a transverse state can be undesirable as it implies that the straight course state along the direction of wave-propagation (the longitudinal state) is unstable. Our hydro-elastic theory suggests that rigid structures are more sensitive to this instability. 

\subsection{Foam mats}
\label{exp}

\begin{figure}[t]
    \centering
  \includegraphics[width = 0.7\linewidth]{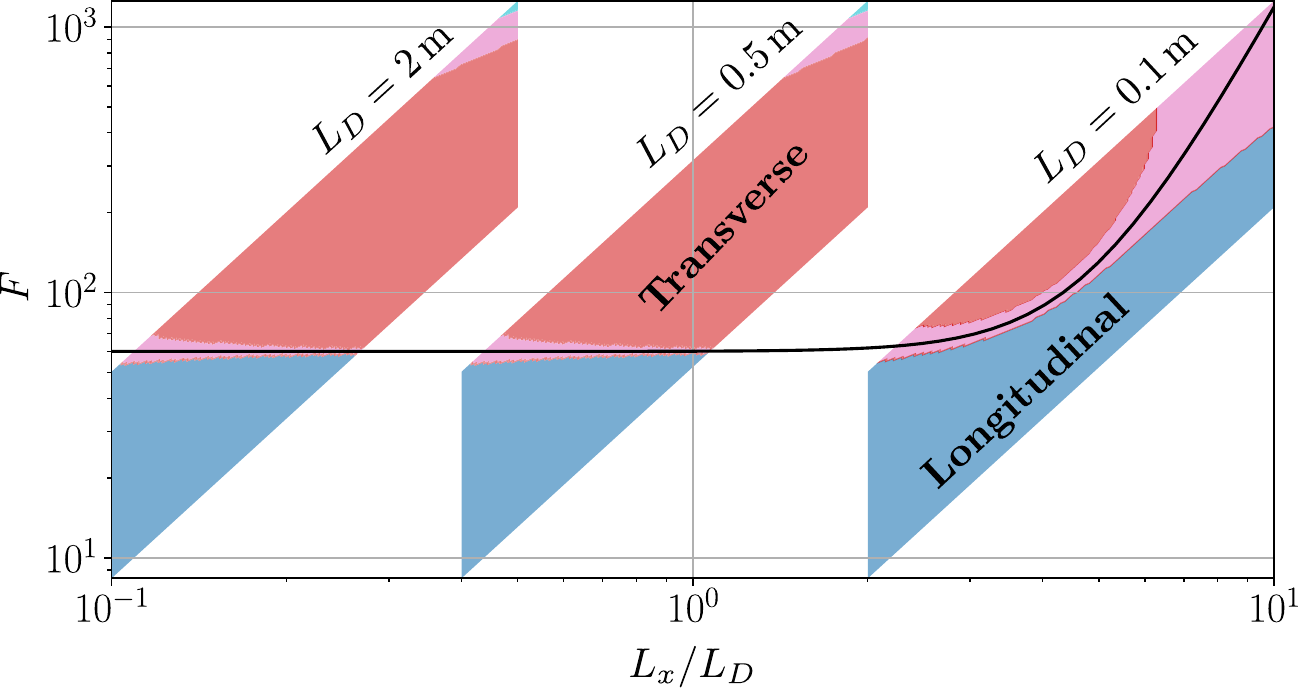}
    \caption{Phase diagram for foam mats with varying flexural lengths, lenghts $L_x \in [0.2,1]$ m and wave lengths  $\lambda \in [0.5,3]$ m.   The black line shows the short floater limit transition line $F =  60 + (5/42)(L_x/L_D)^4$. }
    \label{xpe}
\end{figure}

We finally discuss here the requirements for testing our theoretical predictions in laboratory experiments. To minimize capillary effects, the floaters should have widths of at least several centimeters, which generally corresponds to lengths on the order of one meter. For such structures, medium-scale wave tanks, typically around 30~m in length, supporting wavelengths from one meter up to a few meters, are suitable. As floating material, XPE (cross-linked polyethylene) foam mats, commonly used in swimming pools or aquatic parks, have interesting hydroelastic properties. By adjusting the foam density or the thickness of the mat, the flexural lengths can be varied in the range $L_D \in [0.1,5]$~m.

In figure  \ref{xpe}, we show some predictions on preferential orientation obtained using our theory, for 3 different polyethylene foam mats with flexural lengths $L_D=0.1, 0.5, 2$ m and a fixed draft of $\ovl{h} = 0.01$ m. In each parallelogram-shaped region, we vary the length $L_x \in [0.2,1]$ m and the wave length  $\lambda \in [0.5,3]$~m. This numerical application suggests that rather flexible mats ($L_D \simeq 0.1$~m) are required to observe a significant change in the critical $F_c$.

\section{Conclusion}
\label{sec:conclusion}

Thin and flexible floating structures appear in a wide range of applications, from floating modular pontoons, to floating inflatable structures such as paddleboards or kayaks. Few studies have been dedicated to the second order mean yaw motion of flexible structures in waves. In this study, we have proposed a hydro-elastic theory that can give the second order mean yaw moment on slender deformable structures with arbitrary length and two short directions. Our theory combines the Froude-Krylov approximation (negligible diffraction/radiation) with a Kirchoff-Love model for the bending deformation of the thin pate.  As explained in Appendix \ref{sec:appendix}, diffraction can indeed be neglected for a structure of arbitrary length, when the width and draft are much smaller than the wavelength.

We have studied how the mean yaw moment varies with floater length, draft, wavelength, flexural length and angle of incidence. Using this, we discuss how the mean yaw moment  leads to a preferential orientation in the case of a freely drifting structure. Preferential orientations are defined as the stable angles of incidence for which the mean yaw moment vanishes. 
For floaters that are short with respect to the wavelength in all directions, we find a clear preference for either the longitudinal or a transverse state and  we can precisely locate the transition line. Preferential orientation is controlled by the number $kL_x^2 / \overline{h}$ and how it compares to a critical value $F_c$ that depends on floater shape and the ratio $L_x/ L_D$. Soft, short and heavy floaters prefer the longitudinal state, while stiff, long and light floaters prefer the transverse state. These predictions apply up until floater lengths $L_x  < \lambda/2$. 

For floaters that are longer than half a wavelength, our theory predicts mean yaw moments with a complex variation in angle of incidence. This arises because the pressure forces alternate in sign along the floater, leading to partial cancellation. Long floaters can have intermediate equilibrium orientations, and very long floaters can even have multiple stable equilibria. In that case,  preferential orientation may well be unpredictable.  

We have applied our theory to three practical configurations. For moored, flexible pontoons, provides a means to estimate the mean yaw moment, which can be useful for comparison with numerical solvers. For flexible inflatable floaters such as kayaks or paddle boards, we show that the mean yaw moment can affect the course stability. Finally, we propose an experimental design using polyethylene foam mats in a wave tank, which could quantitatively test our predictions.

Our model can be extended in several directions. An interesting direction is to consider hull shapes more complex than the rectangular parallelepipeds analyzed here, which may be relevant for naval applications. Another is to incorporate capillary effects, which may significantly influence the mean yaw moment and the resulting preferential orientation of floaters at the centimeter scale. This could have implications for the transport of deformable pollutants or drifting sargassum mats. These research directions are left for future work.

\begin{acknowledgments}

We thank A. Eddi, S. Perrard, X. Chen and S. Malenica for fruitful discussions. This work was supported by the project “TransWaves” (Project No. ANR-24-CE51-3840-01) of the French National Research Agency.

\end{acknowledgments}

\begin{appendix}

\section{Why can diffraction be ignored in the limit $L_x \gg \lambda \gg L_y \gg L_z$ ? } 
\label{sec:appendix}

A key assumption in our approach is the Froude-Krylov assumption, which usually requires that the floater is small with respect to the wavelength in all three directions: $\lambda \gg L_x \gg L_y \gg L_z$.  In this Appendix, we explain why a diffractionless theory also applies to slender floaters of arbitrary lengths and only two short dimensions:  $L_x \gg \lambda \gg L_y \gg L_z$. This is based on previous observations in the perfectly flexible and solid limits \cite{herreman2024preferential,dhote2025flexible} that we recall here and extends to the case of flexible structures.

In the case of perfectly flexible thin sheets, diffraction can be safely neglected. This is because flexible sheets perfectly adapt to the instantaneous shape of the wave, keeping both the kinematic and inviscid dynamic boundary conditions identical to those at a free surface. Based on this diffractionless assumption,  \citet{dhote2025flexible}  derived the mean yaw moment for a perfectly flexible slender strip with arbitrary lengths $L_x$ and $\lambda \gg L_y \gg L_z$,
\bse
\ba \label{Kzflex_long}
  \ovl{K}_z^{\,\text{flex}} &=& -  \frac{1}{2} \rho g k^2 a^2   \beta  L_x^2 L_y L_z \sin \overline{\psi}    \left ( \left (  \cos^2 \overline{\psi}  \right )  j_0 (X ) j_1 (X )  + \left (1 - \frac{ \cos^2 \overline{\psi}}{2}  \right ) 
j_1  (X ) j_2   (X ) \right )  
\ea
with $X=  (kL_x/2)  \cos \ovl{\psi}  $,  for brevity. This is equivalent to Eq.~\eqref{Kzlong} for the L-part of the mean yaw moment, in dimensional form.  In the short limit, this formula reduces to 
\be \label{Kzflex_short}
kL_x  \ll 1 \quad : \quad  \ovl{K}_z^{\,\text{flex}} \approx -  \frac{1}{12} \rho g k^3 a^2   \beta  L_x^3 L_y L_z \sin \overline{\psi}    \cos^3 \overline{\psi}  ,
\ee
\ese
which is Eq.~\eqref{Kzsmall_part1} in dimensional form. This mean yaw moment is negative for $\ovl{\psi} \in [0,90^o]$: short flexible floaters drift towards the longitudinal equilibrium, as confirmed by experiments  \cite{dhote2025flexible}.

In the case of solid floaters, diffraction is a priori not negligible for floaters with arbitrary length $L_x$ and $\lambda \gg L_y \gg L_z$.  In his original paper, \citet{newman1967drift} derived an analytical formula for the mean yaw moment on slender and solid rectangular barges. Instead of integrating the pressure force and moment on the moving hull as we do (near-field approach), Newman writes the angular momentum balance on a cylindrical control volume with infinite radius. Using conservation of angular momentum, the mean yaw moment can alternatively be expressed as a surface integral over the cylindrical boundary at infinity. Physically, this integral captures the rate at which mean angular momentum is being radiated away to infinity and it can be calculated with the far-field expansions of the diffracted and radiated waves. This method, now referred to as the far-field approach, necessarily includes both diffraction and radiation; otherwise, no angular momentum can be transported away.
Using a slender body approximation, \citet{newman1967drift} obtains a formula for the mean yaw moment on a solid rectangular barge of arbitrary length, which writes in our notation
\bse \label{newman}
\be \label{newman_long}
\ovl{K}_z^{\, \text{Newman}} =   \frac{1}{2} \rho g k a^2 L_x^2 L_y \sin \overline{\psi}  \,  j_1  \left ( \frac{ kL_x }{2} \cos \ovl{\psi}  \right )   j_2  \left ( \frac{ kL_x }{2} \cos \ovl{\psi}  \ \right ) . 
\ee
Compared to Newman, we define our angle of incidence in the opposite direction. Notice the same type of spherical Bessel functions as in the flexible limit formula \eqref{Kzflex_long}. In the short floater limit, this formula reduces to 
 \be \label{newman_short}
kL_x \ll 1 \quad : \quad   \ovl{K}_z^{\, \text{Newman}} \approx  \frac{1}{720} \rho g k^4 a^2  L_x^5 L_y  \sin \overline{\psi}    \, \cos^3\overline{\psi} .
\ee
\ese
This mean yaw moment is positive in the interval $\ovl{\psi} \in [0,90^o]$ and suggests a transverse preferential orientation for solid, short and slender floaters.

This early prediction of Newman disagreed with our experiments  \cite{herreman2024preferential}, which show that short solid floaters prefer the longitudinal equilibrium.  In Ref.~\cite{herreman2024preferential}, we have reconsidered the calculation of the mean yaw moment on small solid, rectangular barges. Since we only considered floaters small with respect to the wavelength in the experiments, we developed a diffractionless near-field method similar to the one developed in this article, and obtained the mean yaw moment 
\be \label{herreman_short}
kL_x   \ll 1 \quad : \quad   \ovl{K}_z^{\,\text{solid}} \approx  \frac{1}{12} \rho g k^3 a^2  L_x^3 L_y \left ( - \beta L_z + \frac{kL_x^2}{60} \right ) \sin \overline{\psi}    \, \cos^3\overline{\psi} .
\ee
Remarkably, this diffractionless approach yields a moment that is exactly the sum of the flexible limit moment \eqref{Kzflex_short} 
and Newman's moment  \eqref{newman_short}. The term proportional to $-\beta L_z$ relates to the L-part (Eq. \ref{Kzlong}), while the term proportional to $kL_x^2/60$ (Newman's contribution) relates to the T-part (Eq. \ref{Kztrans}), i.e., to first order, to the spatial variation of the submersion depth $h'$. It is this formula that predicts that short floaters with $F < 60$ prefer the longitudinal state and long floaters with $ F > 60$  the transverse state, in agreement with experiments \cite{herreman2024preferential}. Our alternative calculation highlights that Newman's model misses the L-part of the mean moment, related to the first order motion $x_c'$ and $\psi'$. This omission likely stems from an implicit assumption of vanishing draft, which prevents capturing the L-part of the mean yaw moment, proportional to the draft. \citet{Newman_2026} corrected this in a very recent work on preferential orientation in the same short floater or long wavelength limit. Replacing $V_{xx} = \beta L_x^3 L_y L_z / 12$ and taking $T=1$ (infinite depth) in his formula (7.7), this corrected mean yaw moment formula finds the exact same term ($\sim \beta L_z$) that was lacking in the original theory \citet{newman1967drift}. A quantitative comparison to our diffractionless formula is however lacking and would have been useful.

It may seem surprising that our diffractionless model can recover Newman's formula \eqref{newman_short} as a part of the mean yaw moment: diffraction is present in his far-field model whereas it is absent in ours. This calls into question the role of diffraction in these mean yaw moment formula. Diffraction is certainly essential in the far-field method as without it, no mean angular moment can be calculated,  but this does not imply that the near-field pressure is necessarily strongly affected by diffraction. In the near field, the incoming wave pressure can remain dominant and this is certainly what happens in the short floater limit. 

The previous explanation applies to short, slender floaters, but what about slender floaters with arbitrary lengths and only two short directions? In appendix A of Ref.~\cite{herreman2024preferential}, we have used our Froude-Krylov, near-field theory, to calculate the T-part of the mean yaw moment (see Eq. \eqref{Kztrans}) on solid, slender floaters with $\lambda \gg  L_y \gg L_z$ and arbitrary $L_x$. We know that in the short limit, it is this T-part that relates to Newman's formula. This calculation yields
\ba
\ovl{{K}}_z^{T, \text{solid}} &= &  -  \frac{1}{4} \rho g k a^2 L_x^2 L_y  s_\psi  \,  \text{sinc}' \left ( \frac{\ovl{c}_\psi kL_x }{2}  \right )   \left [ \text{sinc}   \left ( \frac{\ovl{c}_\psi kL_x }{2}  \right )  + 3 \, \text{sinc}''  \left ( \frac{\ovl{c}_\psi kL_x }{2}  \right ) \right ]  \nonumber  \\
&= &   \frac{1}{2} \rho g k a^2 L_x^2 L_y \sin \overline{\psi}  \,  j_1  \left ( \frac{ \ovl{c}_\psi kL_x }{2}  \right )  j_2  \left ( \frac{ \ovl{c}_\psi kL_x }{2}  \right )    = 
\ovl{K}_z^{\, \text{Newman}}  
\ea
Using $\text{sinc}' ( X) =- j_1 (X)$ and  $\text{sinc} ( X) + 3\,  \text{sinc}'' ( X) = 2 j_2 (X)$ allows one to recover Newman's formula in our expression. This shows that a diffractionless theory can exactly recover Newman's formula  \eqref{newman_long} for floaters of arbitrary length. This observation suggests that we can rightfully ignore diffraction in the case of long floaters with two short dimensions. Just as in the short limit, it means that the near-field pressure of long slender floaters is almost unaffected by diffraction. 

Since Ref.~\cite{herreman2024preferential}, we have completed the calculation of the total mean yaw moment on solid floaters of arbitrary length, including the L-part that was not captured by Newman. 
We find 
\ba \label{Kzsolid}
 && \ovl{K}_z^{\,\text{solid}} =   \frac{1}{2} \rho g k a^2 L_x^2 L_y \sin \overline{\psi}  \\
 & & \times  \left [  -  \beta k L_z  \left ( \left (  \cos^2 \overline{\psi}  \right )  j_0 (X)   j_1 (X)  + \left (1 - \frac{ \cos^2 \overline{\psi}}{2}  \right ) 
j_1 (X) j_2 (X)  \right ) +   j_1 (X)   j_2  (X)  \right ] \nonumber 
\ea
with, as before, $X = (kL_x /2)  \cos \overline{\psi}$. Just like in the short limit, the total mean yaw moment equals the sum of the L-moment \eqref{Kzflex_long} and the T-moment identified by Newman \eqref{newman_long}.
 The extra L-part of the mean yaw moment on a solid floater is by the way identical to that identified in \eqref{Kzlong} or \eqref{Kzflex_long}, because solid floaters have the same first order motion  $x_c'$ and $\psi'$ as flexible ones. This new formula for solid rectangular barges improves Newman's formula and it should give a decent approximation when $\lambda \gg L_y \gg L_z$. In figure  \ref{Kz_verylong}, we observe that our hydro-elastic, arbitrary length theory reproduces well this solid limit.  

We have discussed the role of diffraction in both very flexible and solid limits. For the intermediate case of elastic floaters, the discussion on the role of diffraction cannot be made as precise, because we cannot compare our theory to other existing results. However, rigid obstacles are more strongly affected by diffraction than flexible structures; therefore, if diffraction is negligible for rigid bodies, it should be even less significant for flexible structures. This justifies why our diffractionless hydro-elastic theory should also apply to floaters of arbitrary length in the limit $\lambda \gg L_y \gg L_z$.

\end{appendix}

\bibliography{apssamp}

\end{document}